\documentclass[12pt]{article}

\usepackage{latexsym,amsmath,amscd,amssymb,framed,graphics,mathrsfs}
\usepackage{framed,esint,slashed}

\usepackage{enumerate}
\usepackage{stackrel}
\usepackage{graphicx}

\usepackage[colorlinks]{hyperref}
\usepackage{url}
\usepackage{colordvi}
\usepackage{color}
\usepackage{bbm}
\usepackage{cancel}

\usepackage{soul}
\setstcolor{blue}
\usepackage{caption}

\usepackage[all]{xy}

\usepackage{frame}

\makeatletter

\@addtoreset{figure}{section}
\def\thefigure{\thesection.\@arabic\c@figure}
\def\fps@figure{h, t}
\@addtoreset{table}{bsection}
\def\thetable{\thesection.\@arabic\c@table}
\def\fps@table{h, t}
\@addtoreset{equation}{section}

\makeatother

\newcommand{\pp}[2]{\frac{\partial #1}{\partial #2}}

\newtheorem{theorem}{Theorem}[section]

\newtheorem{remark}[theorem]{Remark}

\begin{document}
\title{Thermodynamically consistent semi-compressible fluids: a variational perspective}

\author{
  Christopher Eldred\\
  \texttt{celdred@sandia.gov}\\
Org 01446 (Computational Science), Sandia National Laboratories  \and
  Fran\c{c}ois Gay-Balmaz\\
  \texttt{francois.gay-balmaz@lmd.ens.fr 
}\\
  CNRS, Ecole Normale Sup\'erieure de Paris, LMD
}

\date{}

\maketitle

\abstract{
This paper presents (Lagrangian) variational formulations for single and multicomponent semi-compressible fluids with both reversible (entropy-conserving) and irreversible (entropy-generating) processes. Semi-compressible fluids are useful in describing \textit{low-Mach} dynamics, since they are \textit{soundproof}. These models find wide use in many areas of fluid dynamics, including both geophysical and astrophysical fluid dynamics. Specifically, the Boussinesq, anelastic and pseudoincompressible equations are developed through a unified treatment valid for arbitrary Riemannian manifolds, thermodynamic potentials and geopotentials. By design, these formulations obey the 1st and 2nd laws of thermodynamics, ensuring their thermodynamic consistency. This general approach extends and unifies existing work, and helps clarify the thermodynamics of semi-compressible fluids. To further this goal, evolution equations are presented for a wide range of thermodynamic variables: entropy density $s$, specific entropy $\eta$, buoyancy $b$, temperature $T$, potential temperature $\theta$ and a generic entropic variable $\chi$; along with a general definition of buoyancy valid for all three semicompressible models and arbitrary geopotentials. Finally, the elliptic equation is developed for all three equation sets in the case of reversible dynamics, and for the Boussinesq/anelastic equations in the case of irreversible dynamics; and some discussion is given of the difficulty in formulating the elliptic equation for the pseudoincompressible equations with irreversible dynamics.
}


\section{Introduction}

Many important situations in fluid dynamics, especially astrophysical and geophysical fluids, have fluid velocities significant less than the speed of sound. This is known as a \textit{low-Mach} regime, and it motivates the introduction of sound-proof equation sets that do not support sound waves. These equations sets are useful for numerical simulations, and also for improved physical understanding since it isolates the phenomena of interest from the unimportant sound waves.

The most common approach to eliminating sound waves is to modify the compressibility of the fluid, which breaks the density perturbation $\leftrightarrow$ pressure perturbation feedback that drives sound waves. This is the origin of the name \textit{semi-compressible}. Some commonly used equation sets that work in this way are the \textit{Boussinesq} \cite{Bo1903}, \textit{anelastic} \cite{OgPh1962,LiHe1982} and \textit{pseudoincompressible} \cite{Du1989} models. The derivation of such models often proceeds heuristically through scale analysis and the removal of small terms. However, this can easily lead to equation sets that are not dynamically consistent: they lack fundamental conservation principles such as energy or momentum conservation. Additionally, it can be difficult to correctly formulate the thermodynamics in these cases, especially when irreversible processes are introduced. Furthermore, many of these models are developed for specific choices of thermodynamic potentials (usually ideal gas) and thermodynamic variable (usually potential temperature $\theta$), and it is usually unclear how to extend them to more general equations of state or different choices of thermodynamic variable. 

To alleviate these shortcomings, a more robust and general approach is to use a variational formulation: make approximations in the Lagrangian and derive the equations of motion and thermodynamics from this approximated Lagrangian. As long as the approximations do not break the fundamental symmetries of the Langrangian, this ensures that the resulting equation sets will have the standard conservation principles such as energy or momentum. Since there are variational principles for both reversible and irreversible processes, both cases can be consistently handled, and the resulting equations will obey the 1st and 2nd laws of thermodynamics. Additionally, this is a systematic approach that is valid for any thermodynamic potential (equations of state) or geopotential, is coordinate-free and works for general manifolds. 

In the reversible regime and for single component gas, variational derivations of the anelastic and pseudoincompressible have been given in \cite{CoHo2014} for perfect ideal gas and in \cite{VaLeBrWoZw2013} for arbitrary state equations. This approach uses the Euler-Poincar\'e variational formulation, which is induced from the Hamilton principle for fluids via Lagrangian reduction by relabelling symmetry, see \cite{HoMaRa1998}. 

The extension of the Euler-Poincar\'e variational formulations to multicomponent pseudoincompressible models with irreversible processes and with arbitrary state equations was given in \cite{GB2019}. It was used there for the modelling of thermodynamically consistent moist atmospheric dynamics with rain process and subject to the irreversible processes of viscosity, heat conduction, diffusion, and phase transition. This derivation is based on a general variational Lagrangian formalism for nonequilibrium thermodynamics developed in \cite{GBYo2017a,GBYo2017b,GBYo2019} which extends Hamilton?s principle to incorporate irreversible processes.

The inclusion of irreversible processes in the Boussinesq, anelastic, and pseudoincompressible approximations has been traditionally made without addressing the energetic and thermodynamic consistency of the resulting system. Thermodynamically consistent equations for binary fluids and arbitrary state equations were given in \cite{Pa2008} and \cite{Ta2011,Ta2012} for the Boussinesq/anelastic approximation and in \cite{KlPa2012} for the pseudoincompressible approximation. Similar work for the special case of seawater (treated as a binary fluid) for both the anelastic and pseudoincompressible equations are given in \cite{Dewar2015,Dewar2016}.

Extending and unifying previous work, in this paper we will present variational derivations of the single and multicomponent Boussinesq, anelastic and pseudoincompressible equations with irreversible processes for arbitrary thermodynamic potentials (equations of state). Our derivation using variational principles is much simpler, direct, and systematic than previous works and allows for a unified treatment of the multicomponent fully compressible models and these three semi-compressible models. Starting from the Lagrangian of the compressible fluid, given by the kinetic minus the internal and potential energies, the Lagrangian of the pseudoincompressible fluid is obtained by linearization around the background pressure, which immediately produces the pseudoincompressible condition as a constraint in the Lagrangian. The anelastic Lagrangian is then derived by inserting the anelastic condition in the constraint. The general variational formulation of thermodynamics given in \cite{GBYo2017b} applied to these Lagrangians then directly yields the three models, in a form in which thermodynamic consistency is easily achieved. We refer to Figure \ref{figure} for a visual summary. In addition, the variational approach directly yields the needed modification of thermodynamic forces that is required for thermodynamic consistency.

Our approach is completely intrinsic as it does not depend on any coordinate system and is valid for fluid motion on arbitrary Riemannian manifold, which is crucial for the consistent treatment of spherical geometry, and for any geopotential. For simplicity of exposition, we will concentrate on a fluid in $ \Omega \subset \mathbb{R}^3$, but explain how the developments can be extended to arbitrary manifolds in Remark \ref{Remark_Manifolds}. Additionally, we present explicitly the evolution equations for a wide range of thermodynamic variables: entropy density $s$ (from which specific entropy $\eta$ can be easily deduced), buoyancy $b$, temperature $T$, potential temperature $\theta$ and a generic entropic variable $\chi$. We also give a general expression for the resulting elliptic equation in the reversible case, and discuss the difficulties with formulating this equation in the irreversible case. These considerations are useful for practical use of the models.

Our approach also yields a general definition of buoyancy valid for any geopotential $ \phi (x)$. For the single component anelastic ideal gas, we explore different reference pressure profiles and effects on $b$ evolution + other variables.

In doing so, we highlight very strong similarities between the pseudoincompressible and the Boussinesq/anelastic models, especially in the energetics and thermodynamics.

The main assumptions underlying our approach are local thermodynamic equilibrium, and the existing of material boundaries for the manifold. Additionally, in the multicomponent case we assume that the fluid can be characterized with a single (barycentric) velocity and a single temperature.

As we will explain in detail, although the final equations coincide, our treatment of pseudoincompressible models slightly differs from the one in \cite{GB2019} as we use a Lagrangian function which allow us to identify the strong connection between the pseudoincompressible and anelastic models.

The remainder of this paper is structured as follows: Section \ref{sec-2} provides a review of the Lagrangian variational approach for both reversible and irreversible dynamics, in the case of a single component fully compressible fluid. This approach is  applied to single component semi-compressible models in Section \ref{sec-3} and multicomponent semi-compressible models in Section \ref{sec-4}. Finally, in Section \ref{conclusions} some conclusions are drawn. Appendix \ref{appendix_A} provides details of multicomponent pseudoincompressible derivation.

\section{Review of geometric variational setting}
\label{sec-2}

In this section we present a review of the geometric variational formulation of compressible fluids with irreversible processes, which forms the main tool for the developments made in this paper. To keep things simple, we will consider only the case of a single component fluid. The extension to multicomponent fluids is straightfoward and found in \cite{GBYo2017b}.

We first recall the standard situation of \textit{reversible} fluid dynamics. In this case, the equations of motion in the material (or Lagrangian) description follow from Hamilton's principle on groups of diffeomorphisms, associated to the Lagrangian of the fluid model. This geometric description is an extension of the geodesic interpretation of the solutions of the Euler equations on groups of volume preserving diffeomorphisms due to \cite{Ar1966}. We then review the geometric variational formulation deduced in the spatial (or Eulerian) description, given by the Euler-Poincar\'e approach \cite{HoMaRa1998}.
 
In the second part of this section, we review the extension of this geometric variational setting that includes \textit{irreversible processes} in fluid dynamics, such as viscosity, heat condition, or diffusion, \cite{GBYo2017b}. We first describe the approach in the material description, and deduce its Eulerian form, which extends the Euler-Poincar\'e formulation to irreversible processes.

\subsection{Reversible fluid motion}\label{subsec_RFM}

\paragraph{Material description.} In the material description and in absence of irreversible processes, the motion of a fluid in a domain $ \Omega $ is characterized by a time dependent family of diffeomorphisms $ \varphi (t): \Omega \rightarrow \Omega $, so that the trajectory of a fluid particle with label $X \in \Omega $ is found as $x= \varphi (t)(X)= \varphi (t,X)$. For simplicity, we assume in this paper that $ \Omega$ is a domain in $ \mathbb{R} ^n $ with smooth boundary, but all our developments extend to the case where  $ \Omega $ is a Riemannian manifold with smooth boundary, see Remark \ref{Remark_Manifolds}. We shall denote by $ \operatorname{Diff}( \Omega )$ the group of diffeomorphisms of $ \Omega $ and by $ \operatorname{Den}( \Omega )$ the space of densities on $ \Omega $ \footnote{In this paper densities are identified with scalar functions by using the standard volume form on $ \mathbb{R} ^n$.}. The Lagrangians for compressible hydrodynamics are functions of the form
\begin{equation}\label{Lagr_general} 
L=L( \varphi , \dot \varphi , \varrho , S): T \operatorname{Diff}( \Omega ) \times \operatorname{Den}( \Omega ) \times \operatorname{Den}( \Omega ) \rightarrow \mathbb{R},
\end{equation} 
where $\varrho  , S \in \operatorname{Den}( \Omega )$ are the mass density and entropy density of the fluid in the material description. Here $T \operatorname{Diff}( \Omega )$ is the tangent bundle to $ \operatorname{Diff}( \Omega )$ and $( \varphi , \dot \varphi ) \in T \operatorname{Diff}( \Omega )$ refers to an arbitrary element of this tangent bundle, given by a fluid configuration $ \varphi $ and a fluid material velocity $\dot \varphi $.

\medskip 

Hamilton's principle for a compressible fluid with Lagrangian \eqref{Lagr_general} reads
\begin{equation}\label{HP} 
\left. \frac{d}{d\varepsilon}\right|_{\varepsilon=0}  \int_0^T L( \varphi _ \varepsilon  , \dot \varphi _ \varepsilon , \varrho _{0}, S_0){\rm d}t=0,
\end{equation} 
where $ \varphi _ \varepsilon$ is an arbitrary path of time dependent diffeomorphisms prescribed at the temporal extremities $ t=0,T$.
Note that in \eqref{HP}, only the diffeomorphisms are varied, the mass density $ \varrho _0$ and entropy density $S_0$ are held fixed and are time independent, as highlighted by the indices $0$. That $ \varrho _0$ and $S_0$ are constant in time in the material description corresponds to mass and entropy conservation.

\paragraph{Eulerian description.} From the relabelling symmetry of fluid dynamics, the Lagrangian $L$ in \eqref{Lagr_general} must satisfy
\begin{equation}\label{Invariance_L} 
L( \varphi \circ \psi , \dot \varphi \circ \psi , \psi ^* \varrho  , \psi ^* S) = L( \varphi , \dot \varphi , \varrho , S), \quad \forall\; \psi \in \operatorname{Diff}( \Omega ) ,
\end{equation} 
where $ \psi ^*$ denotes the pull-back of a density by $ \psi $, i.e. $\psi ^* \varrho =( \varrho \circ \psi   )J \psi $, with $J \psi  $ the Jacobian of $ \psi $.
From this invariance, $L$ can be written exclusively in terms of Eulerian variables, i.e., we have
\begin{equation}\label{L_ell} 
L( \varphi , \dot \varphi , \varrho , S)=\ell(u, \rho  , s),
\end{equation} 
see \cite{HoMaRa1998}, where $\ell: \mathfrak{X} ( \Omega ) \times \operatorname{Den}( \Omega ) \times \operatorname{Den}( \Omega ) \rightarrow \mathbb{R}$ is the reduced (Eulerian) Lagrangian associated to $L$, with $\mathfrak{X} ( \Omega )$ the space of vector fields on $ \Omega $ tangent to the boundary. Here $u \in \mathfrak{X} ( \Omega )$ is the Eulerian fluid velocity and $ \rho  , s \in \operatorname{Den}( \Omega )$ are the Eulerian mass and entropy densities. These fields are related to their Lagrangian counterpart as
\begin{align*} 
u&= \dot \varphi \circ \varphi ^{-1} \\
\rho &= \varphi _ * \varrho \\
s &=  \varphi _* S.
\end{align*} 

Hamilton's principle \eqref{HP} for $L$ induces the Euler-Poincar\'e variational principle for $\ell$ given by the critical condition
\begin{equation}\label{EP} 
\delta \int_0^T\ell(u, \rho  , s){\rm d}t=0,
\end{equation} 
with respect to variations of the form
\begin{equation}\label{EP_variation}
\begin{aligned} 
\delta u &
= \partial _t \zeta + [u, \zeta ]\\
\delta \rho&
= - \operatorname{div}( \rho  \zeta )\\
\delta s &
= - \operatorname{div}( s \zeta ),
\end{aligned}
\end{equation} 
where $ \zeta \in \mathfrak{X} ( \Omega )$ is an arbitrary time-dependent vector field with $ \zeta |_{t=0,T}=0$, see \cite{HoMaRa1998}. A direct application of the Euler-Poincar\'e principle \eqref{EP}--\eqref{EP_variation} yields the equations of motion
\begin{equation}\label{system_Eulerian} 
\left\{
\begin{array}{l}
\vspace{0.2cm}\displaystyle( \partial _t + \pounds _u) \frac{\delta \ell }{\delta u }= \rho \nabla \frac{\delta \ell}{\delta \rho }+ s \nabla \frac{\delta\ell }{\delta s }\\
\bar D_t \rho =0, \qquad \bar D_t s=0.
\end{array}
\right.
\end{equation} 
In \eqref{system_Eulerian}, $ \delta \ell/ \delta u$, $ \delta \ell/ \delta \rho  $, $ \delta \ell/ \delta s$ denote the functional derivatives of $\ell$ defined as
\[
\left. \frac{d}{d\varepsilon}\right|_{\varepsilon=0} \ell(u+ \varepsilon \delta u, \rho  , s) =\int_ \Omega  \frac{\delta \ell}{\delta u} \cdot \delta u\, {\rm d} x, \quad \left. \frac{d}{d\varepsilon}\right|_{\varepsilon=0} \ell(u, \rho + \varepsilon \delta \rho   , s) =\int_ \Omega  \frac{\delta \ell}{\delta \rho  }   \delta  \rho \, {\rm d} x
\]
similarly for $ \delta \ell/ \delta s$. The Lie derivative operator in \eqref{system_Eulerian} is given by $ \pounds _u m = u \cdot \nabla m + \nabla u^\mathsf{T} \cdot m + m \operatorname{div}u$. We have used the notation
\[
\bar D_t \rho  = \partial _t \rho  + \operatorname{div}( \rho  u), 
\]
for the Lagrangian time derivative of a density.

\paragraph{Fundamental GFD Lagrangian.} In this paper we shall derive all the models, both in the reversible and irreversible cases, by starting from the Lagrangian of a rotating compressible fluid in presence of gravity, given by
\begin{equation}\label{Fund_Lagr} 
\ell(u, \rho  , s) = \int_ \Omega \Big[ \frac{1}{2} \rho  | u| ^2 + \rho  R \cdot u - \rho  e( \alpha  , \eta ) - \rho  \phi \Big]{\rm d} x,
\end{equation}
where $ \eta = s/ \rho  $ is the specific entropy and $\alpha = \frac{1}{\rho}$ is the specific volume.
The first term represents the kinetic energy of the fluid; the second term is the contribution of rotation where the vector field $R$ satisfies $ \operatorname{curl}R=2 \omega $ with $ \omega $ the angular velocity of the fluid domain; the third term denotes the internal energy with $e=e( \alpha  , \eta )$ the state equation of the fluid; the last term is the potential energy with $ \phi $ the geopotential. By using the expressions of the functional derivatives
\[
\frac{\delta \ell}{\delta u} = \rho  u, \qquad \frac{\delta \ell}{\delta \rho  }= \frac{1}{2} | u| ^2 + u \cdot R - e + \alpha  \frac{\partial e}{\partial \alpha  } + \eta \frac{\partial e}{\partial \eta } - \phi , \qquad \frac{\delta \ell}{\delta s}= - \frac{\partial e}{\partial \eta }=-T
\]
in the first equation of \eqref{system_Eulerian} and using $\bar D_t \rho  =0$, one obtains the balance of fluid momentum
\[
\rho  ( \partial _t u + u \cdot \nabla u + 2 \omega \times u)= - \nabla p - \rho  \nabla \phi, \quad \text{with}\quad p = -\frac{\partial e}{\partial \alpha},
\]
see \cite{HoMaRa1998} for the full derivation.

\subsection{Irreversible fluid motion}\label{subsec_IFM}

We shall now quickly describe the extension of the above geometric variational formulation to the case in which the fluid is subject to irreversible processes by following the approach developed in \cite{GBYo2017b}. For simplicity, we only consider here the case of a single component fluid subject to the process of viscosity and heat conduction. The multicomponent case, studied later, is briefly reviewed in \S\ref{subsec_VFMF}, see \cite{GBYo2017b} for details.

\paragraph{Material description.} For heat conducting viscous fluids, besides the Lagrangian, one also needs to specify the phenomenological expressions of the viscous stress tensor and entropy flux, denoted $ P $ and $J_S$ in the material description. The extension of Hamilton's principle to heat conducting fluids given in \cite{GBYo2017b} involves two additional variables, the internal entropy density variable $ \Sigma (t) \in \operatorname{Den}( \Omega )$ and the thermal displacement $ \Gamma (t) \in F( \Omega )$, with $F( \Omega )$ the space of functions on $ \Omega $. This extension of Hamilton's principle imposes two types of constraints to the critical action condition, the phenomenological and variational constraints, related in a systematic way. We shall assume no-slip boundary condition, hence we consider the group $ \operatorname{Diff}_ 0 ( \Omega )$ of diffeomorphisms that keep the boundary $ \partial \Omega $ pointwise fixed.

The extension of Hamilton's principle \eqref{HP} to heat conducting viscous fluids reads as follows: find the curves $\varphi(t) \in \operatorname{Diff}_0( \Omega )$, $ \Sigma (t), S(t) \in \operatorname{Den}( \Omega)$, and $\Gamma (t)\in F( \Omega )$ which are critical for the \textit{variational condition}
\begin{equation}\label{VP_fluid} 
\delta \int_0^T \Big[L\big(\varphi, \dot \varphi,  S, \varrho _0\big) +  \int_ \Omega(S- \Sigma ) \dot \Gamma   {\rm d} X  \Big]{\rm d}t =0
\end{equation}
subject to the \textit{phenomenological constraint}
\begin{equation}\label{KC_fluid}
\frac{\delta  L }{\delta  S}\dot \Sigma= - P: \nabla\dot \varphi  + J_S \cdot \nabla \dot\Gamma  
\end{equation}  
and for variations subject to the \textit{variational constraint}
\begin{equation}\label{VC_fluid}
\frac{\delta   L}{\delta  S}\delta \Sigma = - P: \nabla \delta \varphi  + J _S \cdot \nabla \delta\Gamma 
\end{equation} 
with $\delta \varphi_{t=0,T}= \delta\Gamma_{t=0,T}=0$, and with $\delta\varphi|_{\partial\mathcal{D}}=0$. We refer to \cite{GBYo2017b} for a detailed discussion.

\medskip 

A direct computation shows that the principle \eqref{VP_fluid}--\eqref{VC_fluid} gives the following equations
\begin{equation}\label{NSF_material} 
\left\{
\begin{array}{l}
\vspace{0.2cm}\displaystyle\frac{d}{dt}\frac{ \delta  L }{ \delta  \dot \varphi }   - \frac{\delta  L }{\delta  \varphi }= \operatorname{DIV} P \\
\displaystyle
-\frac{\delta  L }{\delta   S}(\dot S + \operatorname{DIV}J_S ) =P: \nabla\dot \varphi  +J _S \cdot \nabla \frac{\delta  L }{\delta   S},
\end{array}
\right.
\end{equation}
together with the conditions
\[
\dot \Gamma = - \frac{\delta  L}{\delta  S} \quad\text{and}\quad \dot \Sigma = \dot S + \operatorname{DIV}J_S.
\]
Since $- \frac{\delta  L}{\delta  S}$ is identified with the temperature of the system, the first condition obtained from the variations $ \delta S$ imposes that $ \Gamma $ is the thermal displacement. The second condition is associated to $ \delta \Gamma $ and imposes that $ \Sigma $ is the internal entropy density. If the variation $ \delta \Gamma $ is free on $ \partial  \Omega$, it further imposes $J_S \cdot n=0$ on $ \partial \Omega $, i.e., the fluid is adiabatically closed. The first equation in \eqref{NSF_material} gives the Euler-Lagrange equations subject to viscous stress, while the second one is the entropy production equation, both written for a general Lagrangian $L$.

\paragraph{Eulerian description.} The Eulerian versions of  $\dot \varphi , S, \Sigma , \Gamma $ are the Eulerian velocity $u$, Eulerian entropy density $s$, internal entropy density $ \varsigma $, and thermal displacement $ \gamma $ given as
\[
u= \dot \varphi \circ \varphi ^{-1} \in \mathfrak{X} _0( \Omega ) , \quad s= \varphi _* S  \in \operatorname{Den} ( \Omega ), \quad \varsigma = \varphi _* \Sigma \in \operatorname{Den} ( \Omega ), \quad \gamma = \Gamma  \circ \varphi ^{-1} \in F( \Omega ),
\]
where $\mathfrak{X} _0( \Omega )= \{ u \in \mathfrak{X} ( \Omega )\mid u|_ { \partial \Omega }=0\}$ is the space of vector fields on $ \Omega $ vanishing on the boundary.
The Eulerian viscous stress tensor $ \sigma $ and entropy flux $j_s$ are related to its Lagrangian counterparts as
\[
\sigma =   ( P \cdot \nabla \varphi ^\mathsf{T}  ) \circ \varphi ^{-1}  J\varphi^{-1}\qquad\text{and}\qquad   j_s= (\nabla \varphi \cdot J_S) \circ \varphi ^{-1} J \varphi ^{-1}\, \footnote{These two relations are given here in traditional Cartesian formulations. We refer to \cite{GBYo2017b} for the intrinsic differential geometric relation, hence independent on any coordinate systems and valid for $ \Omega $ an arbitrary manifold, see also Remark \ref{Remark_Manifolds}.}.
\]
As before, relabelling symmetries ensure the existence of a reduced Lagrangian $\ell$, see \eqref{L_ell}.

By using these relations, one obtains that the Eulerian version of the principle \eqref{VP_fluid}--\eqref{VC_fluid} reads
\begin{equation}\label{VP_NSF_spatial}
\delta \int_0^T\Big[ \ell(u , \rho , s)+\int_ \Omega (s- \varsigma  )D_t \gamma \, {\rm d} x \Big] {\rm d}t=0,
\end{equation}
together with the reduced phenomenological and variational constraints given by
\begin{equation}\label{KC_NSF_spatial}
\frac{\delta   \ell}{\delta   s} \bar D_t \varsigma   =-  \sigma : \nabla u +  j_s \cdot \nabla D_t \gamma ,
\end{equation} 
\begin{equation}\label{VC_NSF_spatial}
\frac{\delta   \ell}{ \delta  s}  \bar D_\delta  \varsigma   =-\sigma : \nabla  \zeta+ j_s \cdot \nabla D_ \delta \gamma ,
\end{equation}
and the Euler-Poincar\'e constraints
\begin{equation}\label{EP_variation_NSF} 
\delta u = \partial _t \zeta+[ u , \zeta ], \quad \delta \rho =- \operatorname{div}( \rho \zeta).
\end{equation} 
Here $ \zeta \in \mathfrak{X} _0(\Omega )$ and $ \delta  \gamma \in F( \Omega )$ are arbitrary curves with $ \zeta _{t=0,T}=0$ and $ \delta \gamma _{t=0,T}=0$, and $ \delta s, \delta \varsigma \in F( \Omega )$.
In \eqref{VP_NSF_spatial}--\eqref{VC_NSF_spatial} we have used the following notations for the Lagrangian derivatives $D_t$, $\bar D_t$ and variations $D_ \delta $, $\bar D_ \delta $ of functions and densities
\begin{equation}\label{Lagr_derivatives} 
\begin{aligned} 
D_tf &= \partial _t f + u \cdot \nabla f &  & \quad D_\delta  f = \delta f  + \zeta \cdot \nabla f\\
\bar D_tg &= \partial _t g+ \operatorname{div}(gu) & &\quad  \bar D_\delta  g= \delta g  +  \operatorname{div} ( g \zeta ).
\end{aligned}
\end{equation}

Application of \eqref{VP_NSF_spatial}--\eqref{VC_NSF_spatial} yields the following general equations for compressible heat conducting viscous fluid with Lagrangian $\ell$, viscosity process and heat conduction process described by $ \sigma $ and $j_s$:
\begin{equation}\label{NSF_Eulerian} 
\left\{
\begin{array}{l}
\vspace{0.2cm}\displaystyle
( \partial _t + \pounds _ u ) \frac{\delta  \ell}{\delta u }= \rho  \nabla \frac{\delta \ell}{\delta \rho }+ s \nabla \frac{\delta \ell }{\delta  s }+ \operatorname{div}  \sigma \\
\vspace{0.2cm}\displaystyle \bar D_t \rho  =0 \\
\displaystyle - \frac{\delta \ell }{\delta s } (\bar D_ts  + \operatorname{div} j _s) =  \sigma \!: \!\nabla u + j _s\! \cdot  \!\nabla \frac{\delta \ell}{\delta s },
\end{array}
\right.
\end{equation}
with boundary condition $u|_ {\partial \Omega }=0$. The principle also yields the conditions
\begin{equation}\label{additional_conditions} 
D_t \gamma = - \frac{\delta  \ell}{\delta  s}, \qquad \bar D_t \varsigma = \bar D_t s + \operatorname{div} j_s \qquad\text{and}\qquad j_s \cdot n =0 \;\;\text{on}\;\; \partial \Omega,
\end{equation} 
arising from the variations $ \delta s$ and $ \delta \gamma$ exactly as in the material description earlier.
The first condition imposes the variable $ \gamma $ to be the thermal displacement and $ \varsigma $ to be the internal entropy density, both in the Eulerian frame. The last condition arises if $ \delta \gamma $ is arbitrary on $ \partial \Omega $, which implies that the fluid is adiabatically closed. We refer to \cite{GBYo2017b} for the detailed derivation. In absence of irreversible process, i.e. if $ \sigma =0$ and $j_s=0$, system \eqref{NSF_Eulerian} recovers the Euler-Poincar\'e equations \eqref{system_Eulerian}.

\paragraph{Fundamental GFD Lagrangian.} When the Lagrangian \eqref{Fund_Lagr} is chosen, one gets from the system \eqref{NSF_Eulerian} the fluid momentum equation, the mass conservation equation, and the entropy equation
\begin{equation}\label{NSF_GFD_Eulerian} 
\left\{
\begin{array}{l}
\vspace{0.2cm}\displaystyle \rho  ( \partial _t u + u \cdot \nabla u + 2 \omega \times u)= - \nabla p - \rho  \nabla \phi + \operatorname{div} \sigma  \\
\vspace{0.2cm}\displaystyle \bar D_t \rho  =0\\
\displaystyle
T (\bar D_t s + \operatorname{div}j_s) = \sigma : \nabla u - j_s \cdot \nabla T.
\end{array}
\right.
\end{equation}
The temperature and potential temperature equations can be derived from this in a standard way, see, e.g., \cite{GB2019}.

\begin{remark}[Lagrangian VS Eulerian description]{\rm While the Eulerian formulation of fluid dynamics \eqref{NSF_Eulerian} is more commonly used than its Lagrangian counterpart \eqref{NSF_material}, we have first given here the Lagrangian description, since it is in this description that the variational formulation is simpler, namely, an extension of the classical Hamilton principle. The Eulerian variational formulation is then \textit{deduced} from it by the standard process of reduction by symmetry.
}
\end{remark}

\begin{remark}[Intrinsic formulation on manifolds]\label{Remark_Manifolds}{\rm In this paper, for ease of presentation, we have worked with a fluid domain $ \Omega \subset \mathbb{R} ^n $, with canonical inner product and volume form. All our results can be formulated for fluid motion on Riemannian manifolds. In this case, some of the differential operators denoted $\nabla $ earlier  have to be appropriately defined in terms of the Riemannian metric. On a general Riemannian manifold $ \Omega$, with Riemannian metric $g$, the Lagrangian \eqref{Fund_Lagr} reads
\begin{equation}\label{Fund_Lagr_manifold} 
\ell(u, \rho  , s) = \int_ \Omega \Big[ \frac{1}{2} \rho  g(u,u) + \rho  g(R,u) - \rho  e( \alpha  , \eta ) - \rho  \phi \Big] \mu _g,
\end{equation}
where $ \mu _g$ is the Riemannian volume form associated to the metric $g$. System \eqref{NSF_GFD_Eulerian} on Riemannian manifolds reads
\begin{equation}\label{NSF_GFD_Eulerian_manifolds} 
\left\{
\begin{array}{l}
\vspace{0.2cm}\displaystyle
\rho \big(  \partial_t u + u \cdot \nabla u +  2( i _ u \omega   )^\sharp\big) = - \operatorname{grad} p - \rho \operatorname{grad}\phi +\operatorname{div} \sigma \\
\vspace{0.2cm}\displaystyle \bar D_t \rho  =0\\
\displaystyle T (\bar D_t s + \operatorname{div}j _s) = \sigma  ^\flat : \nabla u - j _s \cdot  {\rm d}  T,
\end{array}
\right.
\end{equation}
where the Levi-Civita covariant derivative $ \nabla $, the divergence operator $ \operatorname{div}$ (including the one appearing in $\bar D_t$), and the gradient operator $ \operatorname{grad}$ are all associated to the Riemannian metric $g$. The operator ${\rm d}$, however, is the usual differential of functions. The stress tensor $ \sigma $ is a symmetric 2-contravariant tensor field, $ \flat$ is the lowering index operator associated to the Riemannian metric, and $ \sharp= \flat^{-1} $.
In the balance of momentum, we have $ i _ u \omega  = \omega ( u , \_\,)$, where $ \omega  $ is the 2-form defined in terms of the vector field $ R $ as $2 \omega  = {\rm d}  R^\flat $, with $ {\rm d}$ the exterior derivative. The expression of the equations on Riemannian manifolds is relevant for example for the writing and study of the equations on the spherical Earth. This also plays an important role for the derivation of variational discretization of the equations of geophysical fluid dynamics, where it is crucial to identify the tensorial nature of the various fields and their dependence or not on a given metric, see \cite{GaMuPaMaDe2011,DeGBGaZe2014,BaGB2019a,BaGB2019b,GaGB2020}. See \cite{BrBaBiGBML2019} for application of the variational discretization to the sphere.}
\end{remark}

\section{Variational modelling of single component semi-compressible fluids}
\label{sec-3}

Following the same approach as in Section \ref{sec-2}, we will now derive the single component formulations for several commonly used \textit{semi-compressible} equations: pseudoincompressible, anelastic and Boussinesq. All of these models modify the compressibility of the fluid in some way, hence the name semi-compressible. This modification has the advantage of eliminating sound waves, since it breaks the connection between density perturbations and pressure perturbations that drives them. These equations are widely used in both geophysical and astrophysical fluid dynamics for this reason, since sound waves are often unimportant and the relevant dynamics are in a \textit{low-Mach regime}.

By using a variational approach a general treatment is possible, such that we make no assumptions on the thermodynamic potential (equations of state) or geopotential. Additionally, we do not make any geometric approximations (traditional or shallow-fluid, for example), although it is straightforward to include them. It would also be straightfoward to introduce the (quasi)-hydrostatic assumption as well, by eliminating vertical kinetic energy. However, for simplicity this is not done. From the variational approach all the thermodynamics in \cite{Ta2011} for the Boussinesq and anelastic models and in \cite{KlPa2012} for the pseudoincompressible model are recovered easily. In particular, the partial derivative of $h+ \phi $ with respect to $ \phi $ gives the general expression of the buoyancy in a clear way in terms of the two mass densities $\rho  ( p_0, \eta ) $ and $ \rho  _0$ and for pseudoincompressible, anelastic, and Boussinesq models. This derivation of the pseudoincompressible model is slightly different from the one in \cite{GB2019}, although it leads to the same equations.

\subsection{Enthalpy and Gibbs potential}\label{h_and_g}

As we shall see below, to derive the semi-compressible models, it is advantageous to treat the pressure as an independent variable in the variational formulation. This is achieved by expressing the Lagrangian in terms of the enthalpy $h( p, \eta )$ or the Gibbs potential $g(p,T)$ defined as
\begin{align*} 
h&= e+p\alpha  \\
g& =  e+ p\alpha   - \eta T.
\end{align*} 
As a preparation, we reformulate here the variational formulation for the compressible fluid, in terms of the enthalpy or the Gibbs potential.

\paragraph{Enthalpic description.} In terms of the enthalpy, the Lagrangian \eqref{Fund_Lagr} can be rewritten with the pressure $p$ an independent variable as
\begin{equation}\label{ell_enth} 
\ell_h(u, \rho  , s,p ) = \int_ \Omega \Big[ \frac{1}{2} \rho  | u| ^2 + \rho  R \cdot u - \rho  h( p  , \eta ) +p - \rho  \phi \Big]{\rm d} x.
\end{equation} 
The Euler-Poincar\'e variational formulation \eqref{EP}--\eqref{EP_variation} and its extension with irreversible processes in \eqref{VP_NSF_spatial}--\eqref{EP_variation_NSF} naturally generalize to such Lagrangians by including the criticality condition with respect to arbitrary variations of the additional variable $p$. This results in the system \eqref{system_Eulerian} for the reversible case and the system \eqref{NSF_Eulerian} for the irreversible case, in which $\ell(u, \rho  , s)$ is replaced by $\ell_h(u, \rho  , s, p)$ and with the additional condition
\begin{equation}\label{additional_condition_p} 
0=\frac{\delta \ell_h}{\delta p}= - \rho  \frac{\partial h}{\partial p}( p, \eta ) + 1
\end{equation} 
which imposes the thermodynamic relation $\rho  = \rho  (p, \eta )$. When written in terms of $\ell_h$, \eqref{system_Eulerian} and \eqref{NSF_Eulerian} produce a system of three equations for the four variables $(u, \rho  , s, p)$, which is closed by the additional condition \eqref{additional_condition_p}. It is this variational principle that is used below for semi-compressible models.

\paragraph{Gibbs description.} Proceeding similarly, the Lagrangian \eqref{Fund_Lagr} can be rewritten with pressure $p$ and temperature $T$ as independent variables by using the Gibbs potential as
\begin{equation}\label{ell_gibbs} 
\ell_g(u, \rho  , s,p,T ) = \int_ \Omega \Big[ \frac{1}{2} \rho  | u| ^2 + \rho  R \cdot u - \rho  g( p  , T ) +p - T s  - \rho  \phi \Big]{\rm d} x.
\end{equation} 
Both formulations  \eqref{EP}--\eqref{EP_variation} and \eqref{VP_NSF_spatial}--\eqref{EP_variation_NSF} generalize to such Lagrangians exactly as above while criticality with respect to arbitrary variations of $p$ and $T$ result in the additional conditions
\begin{equation}\label{additional_condition_pT} 
0=\frac{\delta \ell_g}{\delta p}= - \rho  \frac{\partial g}{\partial p}( p, T )   +1 \quad\text{and}\quad 0= \frac{\delta \ell_g}{\delta T}= - \rho  \frac{\partial g}{\partial T}( p, T)   - \eta 
\end{equation} 
which imposes the thermodynamic relations $\rho  = \rho  (p, T )$ and $ \eta = \eta (p,T)$. Systems \eqref{system_Eulerian} and \eqref{NSF_Eulerian} written for $\ell_g$ produce systems of three equations for the five variables $(u, \rho  , s, p)$, which are closed by the additional two conditions \eqref{additional_condition_pT}.

This description can be quite useful in practice, since pressure and temperature are measurable quantities. Therefore, empirical thermodynamic potentials are usually formulated using the Gibbs function. For example, an extremely accurate Gibbs function for a mixture of moist air, liquid water and salt that is used in almost all operational ocean models has been developed in \cite{TEOS2010}.

\subsection{Reference states}

Given a geopotential $\phi(x)$, we consider a hydrostatically balanced, stratified reference configuration with background pressure $ p_0(x)$ 
\begin{equation}\label{rel_rho_0_p_0}
\nabla p_0(x)= -  \rho  _0(x) \nabla \phi(x).
\end{equation}
Note that $\phi$ is an arbitrary function defined on the fluid domain $ \Omega $ and that the considerations here are intrinsic, i.e., independent on a coordinate system and valid for $ \Omega $ a manifold. In Cartesian coordinates $x=(x_1,x_2,x_3)$, a classical example is $\phi(x)= gx_3$, but we do not restrict the development to this case. From \eqref{rel_rho_0_p_0} one deduces that $p_0$ depends on $x$ only through the value $\phi(x)$, i.e., there exists a function $\bar p_0$ such that
\begin{equation}\label{bar_p_0} 
p_0(x)= \bar p_0( \phi(x)),
\end{equation} 
which also implies
\begin{equation}\label{bar_rho_0} 
\rho  _0(x)= \bar \rho  _0( \phi (x))
\end{equation} 
for the background mass density. This fits with the usual thinking about hydrostatic balance, which is assumed to occur in the local vertical direction which is defined by the geopotential.

By considering such a general reference configuration, it is possible to treat both astrophysical and geophysical fluids on arbitrary manifolds with arbitrary geopotentials, greatly extending the utility of the formulations introduced below.

\subsection{Pseudoincompressible model}

\paragraph{Lagrangian for pseudoincompressible models.} We shall linearize the Lagrangian $\ell_h$ in \eqref{ell_enth} around the background pressure  $p_0(x)$. To do this, we first note that for a given pressure $p_0$ and writing $p= p_0+p'$ we have, at first order in pressure variation $p'$, 
\begin{align*} 
\rho  h ( p  , \eta ) - p &=  \rho  h ( p_0+ p'  , \eta ) - p_0 - p'   \\
&\simeq \rho  h ( p_0, \eta ) + \rho  \frac{\partial h}{\partial p}(p_0, \eta ) p'    - p_0 - p'  \\
&=\rho  h ( p_0, \eta ) + \rho   \frac{1}{\rho  (p_0, \eta )} p'   - p_0 - p'  \\
&=\rho  h ( p_0, \eta ) +p'  \left(     \frac{\rho}{\rho  (p_0, \eta )}  - 1 \right)   - p_0 .
\end{align*} 
Using this first order expansion in the Lagrangian \eqref{ell_enth} with respect to the background pressure $p_0(x)$ yields the expression
\begin{equation}\label{ell_PI}
\ell_{\rm pi}(u, \rho  , s, p') = \int_ \Omega \Big[ \frac{1}{2} \rho  | u| ^2 + \rho  R \cdot u  - \rho  h(p_0, \eta )- \rho  \phi  +p' \left( 1 - \frac{\rho  }{ \rho  (p_0, \eta )} \right) + p_0 \Big]{\rm d} x.
\end{equation} 
The pressure perturbation $p'$ appears as a Lagrange multiplier enforcing the constraint $ \rho  = \rho  (p_0(x),\eta ) $. We stress that in the integrand in \eqref{ell_PI}, $p_0$ is an explicit function of $x$, given as $p_0(x)=\bar p_0( \phi (x))$.

\paragraph{Variational derivation of pseudoincompresible thermodynamics.} We apply the variational formulation \eqref{VP_NSF_spatial}--\eqref{EP_variation_NSF}  to the Lagrangian \eqref{ell_PI}. Hence, we have to compute the critical point condition
\begin{equation}\label{VP_NSF_spatial_pi}
\delta \int_0^T\Big[ \ell_{\rm pi}(u , \rho , s,p')+\int_ \Omega (s- \varsigma  )D_t \gamma \, {\rm d} x \Big] {\rm d}t=0,
\end{equation}
subject to the phenomenological and variational constraints given by
\begin{equation}\label{KC_NSF_spatial_pi}
\frac{\delta   \ell_{\rm pi}}{\delta   s} \bar D_t \varsigma   =-  \sigma : \nabla u +  j_s \cdot \nabla D_t \gamma ,
\end{equation} 
\begin{equation}\label{VC_NSF_spatial_pi}
\frac{\delta   \ell_{\rm pi}}{ \delta  s}  \bar D_\delta  \varsigma   =-\sigma : \nabla  \zeta+ j_s \cdot \nabla D_ \delta \gamma ,
\end{equation}
with the Euler-Poincar\'e constraints \eqref{EP_variation_NSF} and arbitrary variations $ \delta p'$. It yields the system
\begin{equation}\label{Eulerian_ell_PI} 
\left\{
\begin{array}{l}
\vspace{0.2cm}\displaystyle
( \partial _t + \pounds _ u ) \frac{\delta  \ell_{\rm pi}}{\delta u }= \rho  \nabla \frac{\delta \ell_{\rm pi}}{\delta \rho }+ s \nabla \frac{\delta \ell _{\rm pi}}{\delta  s }+ \operatorname{div}  \sigma \\
\vspace{0.2cm}\displaystyle \bar D_t \rho  =0, \qquad \frac{\delta \ell_{\rm pi}}{\delta p'}=0  \\
\displaystyle - \frac{\delta \ell_{\rm pi} }{\delta s } (\bar D_ts  + \operatorname{div} j _s) =  \sigma \!: \!\nabla u + j _s\! \cdot  \!\nabla \frac{\delta \ell_{\rm pi}}{\delta s },
\end{array}
\right.
\end{equation}
with boundary condition $u|_ {\partial \Omega }=0$, together with the conditions
\begin{equation}\label{additional_conditions_PI} 
D_t \gamma = - \frac{\delta  \ell_{\rm pi}}{\delta  s}, \qquad \bar D_t \varsigma = \bar D_t s + \operatorname{div} j_s, \qquad\text{and}\qquad j_s \cdot n =0 \;\;\text{on}\;\; \partial \Omega,
\end{equation} 
arising from the variations $ \delta s$ and $ \delta \gamma$ as earlier.
The derivatives of the Lagrangian \eqref{ell_PI} are computed as 
\[
\frac{\delta \ell_{\rm pi}}{\delta u}= \rho  (u+R), \qquad \frac{\delta \ell_{\rm pi}}{\delta s} =  - \frac{\partial h}{\partial \eta } (p_0, \eta )-  p'  \Gamma (p_0, \eta ), \qquad \frac{\delta \ell_{\rm pi}}{\delta p'}= 1 - \frac{\rho  }{ \rho  (p_0, \eta )},
\]
\[
\frac{\delta \ell_{\rm pi}}{\delta \rho}  = \frac{1}{2} | u| ^2 + R \cdot u - h(p_0, \eta ) + \eta \frac{\partial h}{\partial \eta }(p_0, \eta ) - \phi - \frac{p'}{ \rho  ( p_0, \eta )} + p' \eta \Gamma (p_0, \eta ),
\]
where $ \Gamma (p, \eta )= \frac{\partial ^2 h}{\partial p \partial \eta }(p, \eta )$ is the adiabatic temperature gradient. A major insight of this variational derivation is the natural occurrence of the \textit{modified temperature}
\[
T^*:= D_t \gamma = - \frac{\delta \ell_{\rm pi}}{\delta s} =  T(p_0, \eta ) + p'  \Gamma (p_0, \eta ),
\]
given by the temperature associated to the background pressure $p_0(x)$ and $ \eta $, modified by the term $p'  \Gamma (p_0, \eta )$, see the first equation in \eqref{additional_conditions_PI} which follows from the variations $ \delta s$. This is the modified temperature considered in \cite[(27)]{KlPa2012}.
With this, \eqref{Eulerian_ell_PI} yields the system of equations
\begin{equation}\label{Eulerian_PI} 
\left\{
\begin{array}{l}
\vspace{0.2cm}\displaystyle
\rho  ( \partial _t u + u \cdot \nabla u + 2 \omega  \times u) = - ( \rho  - \rho  _0) \nabla \phi - \nabla p' + p' \kappa _{\rm ad}\nabla p_0 +\operatorname{div}  \sigma \\
\vspace{0.2cm}\displaystyle \bar D_t \rho  =0, \qquad \rho  = \rho  (p_0, \eta ) \\
\displaystyle T^*(\bar D_ts  + \operatorname{div} j _s) =  \sigma \!: \!\nabla u - j _s\! \cdot  \!\nabla T^*,
\end{array}
\right.
\end{equation}
where $ \kappa _{\rm ad}= \frac{1}{ \rho  c_s ^2 }= - \rho  \frac{\partial ^2 h}{\partial p ^2 } (p, \eta )$ is the adiabatic compressibility coefficient, evaluated at $(p_0(x), \eta )$. This derivation also uses the relation \eqref{rel_rho_0_p_0}.
System \eqref{Eulerian_PI} form a closed system for $u, \eta , p'$. 
The associated pressure equation will be derived in \S\ref{Pressure_equation}. We refer to Appendix \ref{appendix_A} for details on the derivation of \eqref{Eulerian_PI}.

\begin{remark}[Other form of the Lagrangian and constraint]{\rm In \cite{GB2019}, a different Lagrangian was used to derive the pseudoincompressible model, namely
\begin{equation}\label{Lagr_GB2018} 
\widehat{\ell}(u, \rho  , s, \lambda ) = \int_ \Omega \Big[ \frac{1}{2} \rho  | u| ^2 + \rho  R \cdot u - \rho  e( \rho  , \eta ) - \rho  \phi \Big]{\rm d} x - \int_ \Omega \lambda \big(p( \rho  , \eta ) - p_0 \big){\rm d} x.
\end{equation}
As opposed to \eqref{ell_PI} the first integral is the fundamental GFD Lagrangian \eqref{Fund_Lagr} which doesn't involve any linearization, and $ \lambda $ is a Lagrange multiplier imposing the constraint $p( \rho  , \eta )=p_0$. Note that this Lagrangian is based on the internal energy, rather than the enthalpy, and $p$ is not an independent variable. The Lagrange multiplier is related to $p'$ as
\[
p'= \lambda \rho  c_s ^2 .
\]
By using this relation, one obtains that the pseudoincompressible equations derived in \cite{GB2019} from the Lagrangian \eqref{Lagr_GB2018} are equivalent to the system \eqref{Eulerian_PI}. As we will see below, the use of the Lagrangian \eqref{ell_PI} allows to enlighten the strong connections between the pseudoincompressible equations and the Boussinesq/anelastic equations. It also allows to more explicitly describe the modified thermodynamics of pseudoincompressible models.
}
\end{remark}

\paragraph{Energy conservation.} The total energy density for the fundamental GFD Lagrangian \eqref{Fund_Lagr} is the sum of the kinetic, internal, and potential energy densities
\begin{equation}\label{e_tot} 
e_{\rm tot} = \frac{1}{2} \rho  |u| ^2 + \rho  e( \rho  , \eta  )+ \rho  \phi .
\end{equation} 
The total energy density of the pseudoincompressible model can be obtained in a similar way as the Lagrangian in \eqref{ell_PI}, i.e., by expressing the internal energy density in \eqref{e_tot}  in terms of the enthalpy and by linearizing around the background pressure $p_0(x)$, which gives
\[
e_{\rm pi}= \frac{1}{2} \rho  | u| ^2 + \rho  h(p_0, \eta ) - p _0 + \rho  \phi  - p' \left( 1 - \frac{\rho  }{ \rho  (p_0, \eta )} \right),
\]
with the last term vanishing.
A long, but straightforward computation yields the energy conservation equation
\[
\bar D_t e_{\rm pi}= \operatorname{div} \left(  \sigma \cdot u  - p u - j_s T^*\right)
\]
along the solutions of \eqref{Eulerian_PI}. 
It takes formally the same form as the energy equation for the compressible fluid with total pressure given by $p=p_0+p'$ and in which the modified temperature $T^*$ appears rather than $T$.

The total energy
\begin{equation}\label{E_tot} 
E_{\rm pi}=\int_ \Omega e_{\rm pi}{\rm d}x 
\end{equation} 
of the pseudoincompressible model \eqref{Eulerian_PI}  thus satisfies
\[
\frac{d}{dt} E_{\rm pi}= 0
\]
since $u |_ { \partial \Omega }=0$ and if $j_s \cdot n=0$, i.e., the fluid is adiabatically closed, consistently with the first law of thermodynamics.

\paragraph{Entropy production and phenomenological relations.} The form of the entropy production equation in \eqref{Eulerian_PI} immediately suggests the well-known phenomenological relations
\begin{subequations}\label{friction_stress_NSF}
\begin{align}  
\sigma &=2 \mu  \,(\operatorname{Def} u)+ \left( \zeta -\tfrac{2}{3}\mu \right)(\operatorname{div} u )  \delta \\
T^*j _s&=- k \nabla T^* \hskip 10mm \text{(Fourier law)}
\end{align}
\end{subequations}
with $\operatorname{Def} u= (\nabla u + \nabla  u ^\mathsf{T})/2$ the deformation tensor, and where $ \mu \geq 0 $ is the first coefficient of viscosity (shear viscosity), $ \zeta \geq 0$ is the second coefficient of viscosity (bulk viscosity), and $ k \geq 0$ is the thermal conductivity. The only, although important, difference with the phenomenological relations for compressible fluids is the occurrence of the modified temperature $T^*$ in the Fourier law, which is essential for thermodynamic consistency. As we shall see below, this occurrence of $T^*$ introduces an additional pressure dependence in the elliptic pressure equation.

\paragraph{Definition of buoyancy.} Note that we can write
\[
h( p_0(x), \eta ) + \phi (x)= h\big( \bar p_0(\phi(x)), \eta \big) + \phi (x)
\]
which hence depends on $x$ only through the value $\phi(x)$ of the geopotential. The buoyancy is defined in this general setting as
\begin{equation}\label{def_buoyancy} 
b( \phi  , \eta ):= - \frac{\partial }{\partial \phi } \big( h(\bar p_0( \phi  ), \eta )+ \phi  \big) = \frac{ \bar\rho  _0( \phi  ) -  \rho  (\bar p_0( \phi ), \eta )}{ \rho  (\bar p_0( \phi ), \eta ) } 
\end{equation} 
where we have used \eqref{rel_rho_0_p_0}, \eqref{bar_p_0}, \eqref{bar_rho_0}. From this, the buoyancy satisfies the relation
\[
\nabla \big( h( p_0(x), \eta ) + \phi (x) \big) = - b(\phi(x), \eta ) \nabla \phi.
\]
The definition \eqref{def_buoyancy} extends the one used in \cite{Ta2012} for anelastic models and also recovers the expression of buoyancy (modulo a factor discussed below) proposed by \cite{Pa2008} and \cite{Yo2010}.
 
This expression of buoyancy differs by a factor of $g$ (the gravitational constant as commonly used in geophysical fluids, where $\phi = gz$ is often assumed) from the usual definition of buoyancy. It would be possible to define
\begin{equation}
    \tilde{b} = b \|\nabla \phi\|
\end{equation}
to recover the usual definition of bouyancy in the case $\phi = gz$. However, we prefer the simplicity and generality of our definition, and the strong connection to the thermodynamics of semi-compressible fluids through the introduction of the conjugate pair $(\phi, b)$ that replaces $(\alpha,p)$; which can be defined easily for any coordinate system or choice of $\phi$.

From \eqref{rel_rho_0_p_0} and using several thermodynamic relations we obtain the differential relation
\begin{equation}\label{differential_b} 
{\rm d} b =  \frac{1}{\rho}\big( \rho  _0 ^2\kappa _{\rm ad}   +  {\bar\rho  _0}'( \phi )\big) {\rm d} \phi  +  \rho  _0   \Gamma {\rm d} \eta ,
\end{equation}
where $ \rho  $, $ \kappa _{\rm ad}$, and $ \Gamma $ are all expressed at $(\bar p_0( \phi ), \eta )$. From the entropy equation in \eqref{Eulerian_PI} and \eqref{differential_b}, we get  the buoyancy equation as
\begin{equation}\label{buoyancy_equ_PI} 
D_t b = \frac{1}{\rho  }\big(\rho  _0^2\kappa _{\rm ad}   +  {\bar\rho  _0}'( \phi )\big) D_t\phi +  \frac{\rho  _0 \Gamma }{\rho  T^*} \left(   \sigma : \nabla u - \operatorname{div} ( j _s   T^*) \right).
\end{equation} 
System \eqref{Eulerian_PI} can be expressed for the buoyancy $b$ as a prognostic variable instead of the specific entropy $ \eta $ by replacing the entropy equation of \eqref{Eulerian_PI} with the buoyancy equation \eqref{buoyancy_equ_PI} and using the relation $ \eta = \eta (\phi ,b)$ to rewrite the expressions $ \Gamma (\bar p_0(\phi), \eta )$, $ \kappa _{\rm ad}(\bar p_0(\phi), \eta )$, $T^*( \bar p_0(\phi), \eta )$ and $ \rho  (\bar p_0(\phi), \eta )$ in terms of $(\phi,b)$. In particular, we have
\[
\rho( \phi , b) = \bar\rho  _0( \phi ) \frac{1}{b+1} .
\]
The pseudoincompressible system \eqref{Eulerian_PI} in terms of the buoyancy becomes
\begin{equation}\label{Eulerian_PI_b} 
\left\{
\begin{array}{l}
\vspace{0.2cm}\displaystyle
\partial _t u + u \cdot \nabla u + 2 \omega  \times u =  b \nabla \phi  - \frac{1}{ \rho  } \nabla p' + \frac{p' \kappa _{\rm ad}}{ \rho  }\nabla p_0 +\frac{1}{ \rho  }\operatorname{div}  \sigma \\
\vspace{0.2cm}\displaystyle \bar D_t \rho  =0, \qquad \rho  = \rho  ( \phi , b ) \\
\displaystyle D_t b = \frac{1}{\rho  }\big(\rho  _0^2\kappa _{\rm ad}   +  {\bar\rho  _0}'( \phi )\big) D_t\phi +  \frac{\rho  _0\Gamma }{\rho  T^*} \left(   \sigma : \nabla u - \operatorname{div} ( j _s   T^*) \right).
\end{array}
\right.
\end{equation} 

\paragraph{Temperature and potential temperature formulations.}  To obtain the equation in terms of the temperature, we consider $T$ as a function $T(\phi, \eta )= T(\bar p_0(\phi), \eta )$ and the differential relation
\begin{equation}\label{differential_T} 
{\rm d}T = - \rho  _0 \Gamma {\rm d}\phi + \frac{T}{C_p}{\rm d} \eta,
\end{equation} 
where $ \Gamma $ and the specific heat at constant pressure $C_p$ are evaluated at $(\bar p_0(\phi), \eta )$.
This gives the temperature equation
\begin{equation}\label{temperature_equ_PI} 
D_t T =  - \rho  _0\Gamma D_t \phi  + \frac{T}{ \rho  \, C_p T^*} \left(  
\sigma : \nabla u - \operatorname{div} ( j _s   T^*)\right).
\end{equation}
System \eqref{Eulerian_PI} can be expressed for the temperature $T$ as a prognostic variables instead of the specific entropy $ \eta $ by replacing the entropy equation of \eqref{Eulerian_PI} with temperature equation \eqref{temperature_equ_PI} and using the relation $ T = T ( \phi , \eta )$.

\medskip 

To obtain the equation in terms of the potential temperature defined by $ \theta ( \eta )= T(p_{00}, \eta )$ for some reference value $p_{00}$ of the pressure, we use the differential relation
\begin{equation}\label{differential_theta} 
{\rm d} \theta = \frac{\theta}{C_p(p_{00}, \eta  ) }{\rm d} \eta  
\end{equation} 
where $C_p$ is evaluated at the pressure $p_{00}$. This gives
\[
D_ t\theta = \frac{ \theta }{ \rho  C_p(p_{00}, \theta  )  T^*} ( \sigma : \nabla u - \operatorname{div}(j_sT^*)). 
\]
and allows to express system \eqref{Eulerian_PI} in terms of $ \theta $ instead of $ \eta $.

\paragraph{General entropic variable.}
Finally, let us consider a general \textit{entropic variable} $\chi = \chi(\eta)$ which is a function of specific entropy. This has the differential relation
\begin{equation}
{\rm d}\chi = \pp{\chi}{\eta}{\rm d} \eta,
\end{equation} 
which gives
\[
D_ t\chi = \pp{\chi}{\eta} \frac{1}{ \rho  T^*} ( \sigma : \nabla u - \operatorname{div}(j_sT^*)). 
\]
In fact, the potential temperature $\theta$ discussed above is one such example of an entropic variable. In ocean models, another important one is the \textit{potential enthalpy} which is widely used under the name \textit{conservative temperature}.

\paragraph{Formulation using the Gibbs potential.} We can also start from the fundamental GFD Lagrangian written in terms of the Gibbs potential \eqref{ell_gibbs}, rather than the enthalpy as in \eqref{ell_enth}, which is useful in practice since empirical thermodynamic potentials are usually formulated using the Gibbs function. Linearization around the reference pressure $p_0$ yields
\[
\ell_{\rm pi}(u, \rho  , s, p', T) = \int_ \Omega \Big[ \frac{1}{2} \rho  | u| ^2 + \rho  R \cdot u  - \rho  g(p_0,T )- \rho  \phi  -sT +p' \left( 1 - \frac{\rho  }{ \rho  (p_0, T )} \right) +p_0 \Big]{\rm d} x
\]
instead or \eqref{ell_PI}. By applying the variational principle \eqref{VP_NSF_spatial}--\eqref{EP_variation_NSF}  to this Lagrangian, with arbitrary variations $ \delta p'$ and $ \delta T$ we get the pseudoincompressible thermodynamic model derived above, with $T$ the prognostic variable in place of $ \eta $. The details are left to the interested reader. In particular, the buoyancy is defined as above by
\[
b( \phi  , T ):= - \frac{\partial }{\partial \phi } \big( g(\bar p_0( \phi  ), T )+ \phi  \big) = \frac{ \bar\rho  _0( \phi  ) -  \rho  (\bar p_0( \phi ), T )}{ \rho  (\bar p_0( \phi ), T ) } , 
\]
where  $\rho = \rho  (\bar p_0( \phi ), T )$ is now written in terms of the temperature rather than the entropy.

\subsection{Anelastic and Boussinesq models}\label{ABM}

\paragraph{Lagrangian for anelastic models.}
The anelastic and Boussinesq approximations are obtained by imposing $ \rho  $ to be equal to the stratification mass density profile $ \rho  _0(x)$ in  \eqref{rel_rho_0_p_0}  rather than to the mass density $ \rho  = \rho  ( \bar p_0( \phi ),  \eta )$ associated to $p_0$ and the entropy $ \eta $. We thus replace $p'  \left(   1-   \frac{\rho}{\rho  (\bar p_0( \phi ) , \eta )}   \right)$ by $p'  \left(   1-   \frac{\rho}{\rho _0}   \right)$ in the Lagrangian \eqref{ell_PI}, thereby giving the Lagrangian
\begin{equation}\label{ell_an}
\ell_{\rm an}(u, \rho  , s, p') = \int_ \Omega \Big[ \frac{1}{2} \rho  | u| ^2 + \rho  R \cdot u  - \rho  h(p_0, \eta )- \rho  \phi  +p' \left( 1 - \frac{\rho  }{ \rho_0} \right) + p _0  \Big]{\rm d} x.
\end{equation} 
To obtain the Boussinesq approximation, we further assume that $\rho_0(x) = \rho_0$, a constant independent of $x$.

\paragraph{Variational derivation of anelastic thermodynamics.} We apply the variational formulation \eqref{VP_NSF_spatial}--\eqref{EP_variation_NSF}  to the Lagrangian \eqref{ell_an}. Hence, we compute the critical point condition
\begin{equation}\label{VP_NSF_spatial_an}
\delta \int_0^T\Big[ \ell_{\rm an}(u , \rho , s,p')+\int_ \Omega (s- \varsigma  )D_t \gamma \, {\rm d} x \Big] {\rm d}t=0,
\end{equation}
subject to the phenomenological and variational constraints given by
\begin{equation}\label{KC_NSF_spatial_an}
\frac{\delta   \ell_{\rm an}}{\delta   s} \bar D_t \varsigma   =-  \sigma : \nabla u +  j_s \cdot \nabla D_t \gamma ,
\end{equation} 
\begin{equation}\label{VC_NSF_spatial_an}
\frac{\delta   \ell_{\rm an}}{ \delta  s}  \bar D_\delta  \varsigma   =-\sigma : \nabla  \zeta+ j_s \cdot \nabla D_ \delta \gamma ,
\end{equation}
with the Euler-Poincar\'e constraints \eqref{EP_variation_NSF} and arbitrary variations $ \delta p'$. It yields the system
\begin{equation}\label{Eulerian_ell_an} 
\left\{
\begin{array}{l}
\vspace{0.2cm}\displaystyle
( \partial _t + \pounds _ u ) \frac{\delta  \ell_{\rm an}}{\delta u }= \rho  \nabla \frac{\delta \ell_{\rm an}}{\delta \rho }+ s \nabla \frac{\delta \ell _{\rm an}}{\delta  s }+ \operatorname{div}  \sigma \\
\vspace{0.2cm}\displaystyle \bar D_t \rho  =0, \qquad \frac{\delta \ell_{\rm an}}{\delta p'}=0  \\
\displaystyle - \frac{\delta \ell_{\rm an} }{\delta s } (\bar D_ts  + \operatorname{div} j _s) =  \sigma \!: \!\nabla u + j _s\! \cdot  \!\nabla \frac{\delta \ell_{\rm an}}{\delta s },
\end{array}
\right.
\end{equation}
with boundary condition $u|_ {\partial \Omega }=0$, together with the conditions \eqref{additional_conditions_PI} for $\ell_{\rm an}$. The functional derivatives of the Lagrangian \eqref{ell_an} are computed as 
\[
\frac{\delta \ell_{\rm an}}{\delta u}= \rho  (u+R), \qquad \frac{\delta \ell_{\rm an}}{\delta s} =  - \frac{\partial h}{\partial \eta } (p_0, \eta ), \qquad \frac{\delta \ell_{\rm an}}{\delta p'}= 1 - \frac{\rho  }{ \rho_0}
\]
\[
\frac{\delta \ell_{\rm an}}{\delta \rho}  = \frac{1}{2} | u| ^2 + R \cdot u - h(p_0, \eta ) + \eta \frac{\partial h}{\partial \eta }(p_0, \eta ) - \phi - \frac{p'}{ \rho _0}.
\]
With this, \eqref{Eulerian_ell_an} yields the system
\begin{equation}\label{BA_intermediate} 
\left\{
\begin{array}{l}
\vspace{0.2cm}\displaystyle
 \partial _t  u + u \cdot\nabla  u + 2 \omega \times u = - \nabla \frac{p'}{\rho_0} + b \nabla \phi  + \frac{1}{\rho}\operatorname{div}\sigma\\
\vspace{0.2cm}\displaystyle  T (\bar D_t s + \operatorname{div} j_s )  = \sigma  : \nabla u- j_s \cdot  \nabla T\\
\displaystyle \bar D_t \rho  =0, \qquad \rho  = \rho  _0,
\end{array}
\right.
\end{equation} 
where $T=T( p_0, \eta )$ and $b( \phi , \eta )$ is the buoyancy defined in \eqref{def_buoyancy}. Using the last equation, the system becomes
\begin{equation}\label{BA_final} 
\left\{
\begin{array}{l}
\vspace{0.2cm}\displaystyle
 \partial _t  u + u \cdot\nabla u + 2 \omega \times u = - \nabla \frac{p'}{\rho_0} + b \nabla \phi + \frac{1}{\rho_0}\operatorname{div}  \sigma \\
\vspace{0.2cm}\displaystyle  T ( \rho  _0 D_t \eta  + \operatorname{div} j _s ) = \sigma : \nabla u -j _s \cdot  \nabla T\\
\vspace{0.2cm}\operatorname{div}( \rho_0  u)=0.
\end{array}
\right.
\end{equation}
Note that with the definition of buoyancy made above, the momentum equation of the anelastic system \eqref{BA_final} with arbitrary state equations, takes the same form as the usual Boussinesq equation, see \cite{Ta2011,Ta2012}. Note also that the condition $\operatorname{div}( \rho_0  u)=0$ is always true for anelastic models, with or without irreversible processes, and for any state equations. This is in sharp constrast with the condition $\operatorname{div}( \rho_0  \theta _0 u)=0$ for pseudoincompressible models, which only holds for perfect gas and in absence of irreversibility.

\paragraph{Buoyancy formulation.} From the form of the balance of fluid momentum in \eqref{BA_final}, it is advantageous to rewrite system \eqref{BA_final} in terms of the buoyancy rather than the entropy. The differential of the buoyancy is
\begin{equation}\label{differential_b_an} 
{\rm d} b = \frac{1}{\rho  }\big(\rho  _0^2\kappa _{\rm ad}   +  {\bar\rho  _0}'( \phi )\big) {\rm d}\phi +  \rho  _0 \Gamma {\rm d} \eta ,
\end{equation} 
where $ \rho  $, $ \kappa _{\rm ad}$, and $ \Gamma $ are all expressed at $(\bar p_0( \phi ), \eta )$ (in particular $ \rho $ in \eqref{differential_b_an} is not equal to $ \rho  _0$). From the entropy equation in \eqref{BA_final}, we get  the buoyancy equation as
\begin{equation}\label{buoyancy_equ_an} 
D_t b = \frac{1}{\rho  }\big(\rho  _0^2\kappa _{\rm ad}   +  {\bar\rho  _0}'( \phi )\big) D_t \phi  +  \frac{\Gamma }{  T} \left(   \sigma : \nabla u - \operatorname{div} ( j _s   T) \right).
\end{equation} 
System \eqref{BA_final} can be expressed for $b$ as a prognostic variable instead of $ \eta $ by replacing the entropy equation of \eqref{BA_final} with \eqref{buoyancy_equ_an} and using the relation $ \eta = \eta ( \phi ,b)$ to rewrite the expressions $ \Gamma (\bar p_0( \phi ), \eta )$, $ \kappa _{\rm ad}(\bar p_0( \phi ), \eta )$, and $ \rho  (\bar p_0( \phi ), \eta )$ in terms of $( \phi ,b)$. In particular, the mass density $ \rho  $ in \eqref{buoyancy_equ_an} is
\[
\rho( \phi ,b) = \bar\rho  _0( \phi ) \frac{1}{b+1} .
\]
Note how \eqref{buoyancy_equ_an} differs from its pseudoincompressible couterpart \eqref{buoyancy_equ_PI}.

\paragraph{Temperature and potential temperature formulations.} By using the differential relation \eqref{differential_T} for the temperature $T(\bar p_0(\phi),\eta )$, we can rewrite the entropy equation in  \eqref{BA_final} as
\begin{equation}\label{temperature_equ_an} 
D_t T =  - \rho  _0\Gamma D_t \phi  + \frac{1}{ \rho_0 C_p } \left(  
\sigma : \nabla u - \operatorname{div} ( j _s   T)\right),
\end{equation}
which allows to express system \eqref{BA_final} with $T$ as a prognostic variable instead of $ \eta $.

\medskip 

Similarly, by using the differential relation \eqref{differential_theta} for the potential temperature $ \theta ( \eta )= T( p_{00}, \eta )$, we can rewrite the entropy equation in  \eqref{BA_final} as
\[
D_ t\theta = \frac{ \theta }{ \rho_0  C_p(p_{00}, \theta  )  T} ( \sigma : \nabla u - \operatorname{div}(j_sT))
\]
which allows to express system \eqref{BA_final} with $ \theta $ as a prognostic variable instead of $ \eta $. Note that the momentum equation doesn't take a particularly simple or remarkable form when expressed in terms of the potential temperature for arbitrary state equations. We shall see in the next subsection under which conditions the momentum equation reduces to its traditional expression in terms of $ \theta $ for perfect gas. See also Remark \ref{GSE} for the general case.

\paragraph{General entropic variable.}
Similarly to the pseudoincompressible case, the evolution equation for a general entropic variable $\chi$ is
\[
D_ t\chi = \pp{\chi}{\eta} \frac{1}{ \rho  _0 T} ( \sigma : \nabla u - \operatorname{div}(j_sT)). 
\]

\subsection{Perfect gas and specific pressure profiles}

Up until this point we have been completely general in terms of the thermodynamic potential and the pressure reference profile $\bar p_0( \phi )$. However, most of the existing semi-compressible models in the literature make specific choices for these, especially the many flavors of anelastic equations. Therefore in this section we will consider the consequences of assuming a perfect ideal gas along with various choices of pressure reference profile for the anelastic equations. In particular, it will be shown that certain choices lead to simplifications in the expression of buoyancy in terms of the predicted thermodynamic variable (such as $\theta$ or $T$).

\paragraph{Perfect gas.} Consider a perfect gas, i.e. $p \alpha= RT$ with $R$ the specific gas constant, and with constant heat capacities $C_v$ and $C_p$. System \eqref{BA_final} as well as its various formulations in terms of either $b$, $T$, or $ \theta $, are obtained by using the expressions of the coefficients $ \kappa _{ad}$ and $ \Gamma $ for perfect gas, as well as the expression of the buoyancy $b$ in terms of the chosen prognostic variable. For instance, denoting constant reference values with the index $00$, the following explicit expressions of $b$ and $T$ are useful.
\begin{itemize}
\item In the system \eqref{BA_final} written for $ \eta $ we use the relations
\begin{align}
b( \phi , \eta )&=  \frac{ \bar\rho  _0( \phi )}{ \rho  _{00}}  \left( \frac{p_{00}}{\bar p_0( \phi )} \right) ^{C_v/C_p} e^{ \frac{1}{C_p}( \eta - \eta _{00})}-1\label{b_Z_eta}\\
T( \phi , \eta )&= T_{00} \left( \frac{\bar p_0( \phi )}{p_{00}} \right)  ^{ R/ C_p} e^{ \frac{1}{C_p}( \eta - \eta _{00})}.\label{T_Z_eta}
\end{align}
\item For the system written in terms of $b$ we use the relation
\begin{equation}\label{T_Z_b}
T( \phi , b)= \frac{\bar p_0(\phi ) }{\bar\rho  _0( \phi ) R}(b+1).
\end{equation}
\item For the system written in terms of $T$ we use the relation
\begin{equation}\label{b_T}
b(\phi , T)= \frac{\bar \rho  _0( \phi ) R T}{\bar p_0( \phi )} -1  .
\end{equation} 
\item For the system written in terms of $\theta$ we use the relations
\begin{align}
b( \phi , \theta ) &=  \frac{ \bar\rho  _0( \phi )}{ \rho  _{00}} \frac{\theta }{\theta _{00}} \left( \frac{p_{00}}{\bar p_0( \phi )} \right) ^{C_v/C_p}  -1\label{b_Z_theta} \\
T( \phi , \theta )&= \theta \left( \frac{\bar p_0( \phi )}{p_{00}} \right)  ^{ R/ C_p}.\label{T_Z_theta}
\end{align}
 
\end{itemize}

\paragraph{Pressure profile 1: $b \sim T$.} We consider the special case in which the buoyancy is independent of $ \phi $ and proportional to the temperature:
\begin{equation}\label{profile_1} 
b( \phi ,T) =b(T)= \alpha_{00} (T-T_{00}),
\end{equation} 
with $ \alpha_{00} = \frac{1}{T_{00}}$. In this case, from the expression of the buoyancy \eqref{b_T} we get a pressure profile of the form
\[
\bar p_0( \phi )= \bar p_0 ( \phi _0)e^{-( \phi - \phi _0)/RT_{00}}=  \bar p_0( \phi _0)e^{ - \rho  _{00}( \phi - \phi _0)/ p_{00}}.
\]
for some constant $ \phi _0$. The associated density profile is
\[
\bar\rho  _0( \phi )= \frac{\bar p_0( \phi _0)}{RT_{00}}e^{-( \phi - \phi _0)/RT_{00}}= \rho  _{00} \frac{\bar p_0( \phi_0 )}{p_{00}} e^{-( \phi - \phi _0)/RT_{00}}.
\]
In this case, the most appropriate descriptions are those using $b$ or $T$ as prognostic variables.

\medskip

The explicit form of the anelastic system with the temperature $T$ as prognostic variable (assuming an ideal gas and the pressure profile above) is
\begin{equation}\label{BA_final_T_1}
\left\{
\begin{array}{l}
\vspace{0.2cm}\displaystyle
\partial _t  u + u \cdot\nabla u + 2 \omega \times u = - \nabla \frac{p'}{\rho_0} + \alpha_{00} (T-T_{00})\nabla \phi + \frac{1}{\rho_0}\operatorname{div}  \sigma\\
\vspace{0.2cm}\displaystyle  D_t T  = - \frac{T}{  C_p T_{00}} D_t \phi  + \frac{1}{ C_p\rho_0 } \left(  
\sigma : \nabla u - \operatorname{div} ( j _s   T)\right) \\
\vspace{0.2cm}\operatorname{div}( \rho_0 u )=0,
\end{array}
\right.
\end{equation} 
which follows from \eqref{temperature_equ_an} for perfect gas and from \eqref{profile_1}. Recall that $ \rho  _0= \bar\rho  _0( \phi )$ and $p_0=\bar p_0( \phi )$ are functions of $ \phi $ while $T_{00}$ is a constant.

\medskip 

In terms of the buoyancy $b$ one obtains from the relation \eqref{profile_1} 

\begin{equation}\label{BA_final_b_1}
\left\{
\begin{array}{l}
\vspace{0.2cm}\displaystyle
\partial _t  u + u \cdot\nabla u + 2 \omega \times u = - \nabla \frac{p'}{\rho_0} +b \nabla \phi + \frac{1}{\rho_0}\operatorname{div}  \sigma\\
\vspace{0.2cm}\displaystyle  D_t b = - \frac{1}{C_p T_{00}}(b+1)D_t \phi  +  \frac{1}{C_p T_{00}\rho  _0 }\left(  
\sigma : \nabla u - \operatorname{div} ( j _s   T)\right)  \\
\vspace{0.2cm}\operatorname{div}( \rho_0 u )=0
\end{array}
\right.
\end{equation}

\paragraph{Pressure profile 2: $b \sim \theta $.} We consider now the case in which the buoyancy is independent of $ \phi $ and proportional to the potential temperature:
\begin{equation}\label{profile_2_b} 
b( \phi , \theta ) =b( \theta ) = \alpha_ {00}  ( \theta - \theta _{00}),
\end{equation}
with $ \alpha_ {00}  = \frac{1}{ \theta _{00}}$. In this case, from \eqref{b_Z_theta} we get a pressure profile of the form
\[
\bar p_0( \phi )= p_{00}\left( C  -  \frac{ \phi - \phi _0}{C_p  \theta _{00}}    \right) ^{C_p/R},
\]
with $C=(\bar p_0( \phi _0)/p_{00})^{R/C_p}$. 
The associated density profile
\begin{equation}\label{Profile_2} 
\bar\rho  _0( \phi ) = \rho  _{00} \left( C - \frac{ \phi - \phi _0}{C_p \theta _{00}} \right) ^{C_v/R}. 
\end{equation} 
In this case, the most appropriate descriptions are those using $b$ or $ \theta $ as prognostic variables.

\medskip 

The explicit form of the anelastic system with the potential temperature $\theta$ as prognostic variable (assuming an ideal gas and the pressure profile above) is
\begin{equation}\label{BA_final_theta_2} 
\left\{
\begin{array}{l}
\vspace{0.2cm}\displaystyle
\partial _t  u + u \cdot\nabla u + 2 \omega \times u = - \nabla \frac{p'}{\rho_0} + \alpha_{00}( \theta - \theta _{00}) \nabla \phi + \frac{1}{\rho_0}\operatorname{div}  \sigma\\
\vspace{0.2cm}\displaystyle D_t \theta=  \frac{  R\theta _{00}}{C_p \bar p_0( \phi )} ( \sigma : \nabla u - \operatorname{div}(j_sT))\\
\vspace{0.2cm}\operatorname{div}( \rho_0  u )=0,
\end{array}
\right.
\end{equation} 
where we have used
\[
\frac{\theta }{C_pT \rho  _0}= \frac{1 }{C_p\rho  _0 } \left( 1- \frac{ \phi - \phi _0}{C_p \theta _{00}} \right) ^{-1}   = \frac{  R\theta _{00}}{C_p \bar p_0( \phi )} 
\]
which follows from \eqref{T_Z_theta} and \eqref{Profile_2}. In absence of irreversible processes ($ \sigma =0$, $j_s=0$), system \eqref{BA_final_theta_2} recovers the usual anelastic equations, \cite{LiHe1982}. We would like to stress that the expression of the momentum equation in terms of $ \theta $ is not well suited for its generalization to general state equations and irreversible processes as seen in \S\ref{ABM}.

\medskip 

By using \eqref{profile_2_b}, we get the system in terms of the buoyancy as
\begin{equation}\label{BA_final_b_2} 
\left\{
\begin{array}{l}
\vspace{0.2cm}\displaystyle
\partial _t  u + u \cdot\nabla u + 2 \omega \times u = - \nabla \frac{p'}{\rho_0} + b \nabla \phi  + \frac{1}{\rho_0}\operatorname{div}  \sigma \\
\vspace{0.2cm}\displaystyle D_t b= \frac{R}{C_p \bar p_0( \phi )} ( \sigma : \nabla u - \operatorname{div}(j_sT))\\
\vspace{0.2cm}\operatorname{div}( \rho_0 u )=0.
\end{array}
\right.
\end{equation}

\begin{remark}[Reversible case]{\rm While both cases have $D_t \theta =0$ in the reversible case (as it is the case for  general state equations and general pressure profile $p_0$), only case 2 has $D_tb=0$ in the reversible case, see \eqref{BA_final_b_1} and \eqref{BA_final_b_2}. This is due to the relation \eqref{profile_2_b} in which $b$ only depends on $ \theta $ and not on $ \phi $.}
\end{remark}

\begin{remark}[General state equations and pseudoincompressible equations]\label{GSE}{\rm The analysis developed here easily extends to general state equations, to more general relations than \eqref{profile_1} and \eqref{profile_2_b} and to the pseudoincompressible equations. The corresponding pressure profile can be determined following the same approach: given a state equation and a desired relation such as $b=b( \theta )$, one gets from \eqref{def_buoyancy} a differential equation to be solved for the pressure profile $p_0$.
}
\end{remark}

\subsection{Diagram}

The link between fully compressible, pseudoincompressible, and anelastic models for both the reversible and irreversible cases is summarized from a variational point of view in the following diagram.


\begin{center}
\includegraphics[width=15cm]{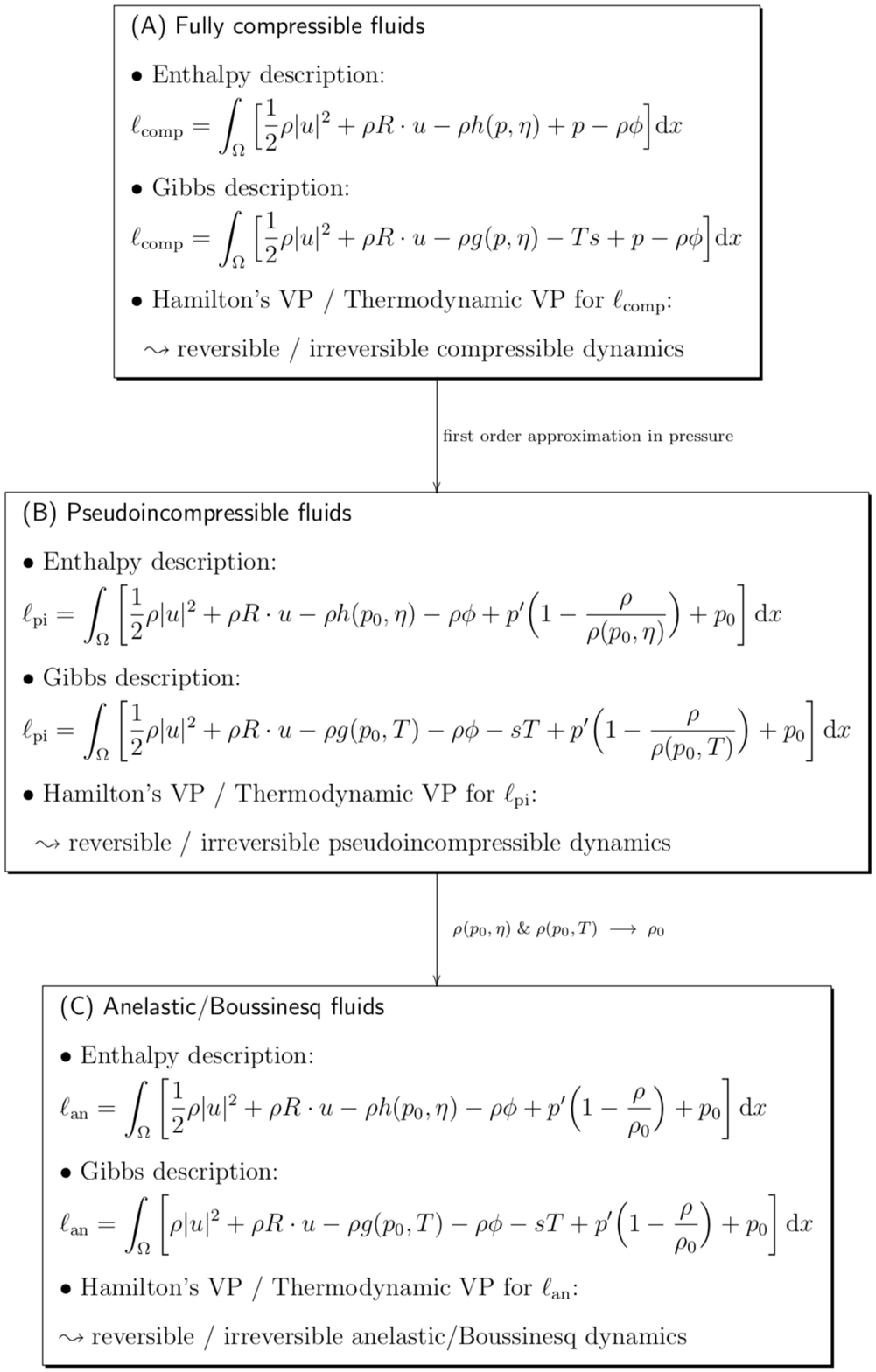}
\end{center}

\noindent Diagram: summary of the variational formulation of compressible and semi-compressible thermodynamics (VP = variational principle).

\subsection{Elliptic pressure equation}\label{Pressure_equation}

To close the equation sets developed above, an equation to determine the Lagrange multiplier $p^\prime$ must be determined. Recall that $p^\prime$ enforces the pseudoincompressible or anelastic constraint, denoted generically $\mathcal{C}=0$. The equation for $p^\prime$ can be derived using standard techniques from variational methods, by discovering the \textit{hidden constraints} associated with $\mathcal{C}$. This is described below.

\subsubsection{Reversible Dynamics}\label{subsec_reversible_dynamics}
Consider the general GFD Lagrangian in enthalpy form for semi-compressible fluids, which is
\begin{equation}\label{general_ell_GFD}
\ell(u,\rho,s,p^\prime) = \widetilde{\ell}(u,\rho,s) + \mathcal{C}(\rho,s,p')
\end{equation}
with
\begin{equation}\label{Constraint_C}
\mathcal{C}( \rho  , s, p') = \int_ \Omega  p^\prime \left(1 - \frac{\rho}{F(\phi,\eta)}\right){\rm d}x
\end{equation}
for some arbitrary function $F(\phi,\eta)$.

The functional derivatives are
\begin{equation}
\frac{\delta \ell}{\delta u  }= \frac{\delta \tilde{\ell}}{\delta u}, \qquad\frac{\delta  \ell}{\delta \rho  } = \frac{\delta \tilde{\ell}}{\delta \rho  } - \frac{p^\prime}{F} - \frac{p^\prime \eta}{F^2} \pp{F}{\eta}, \qquad  \frac{\delta \ell}{\delta s } = \frac{\delta \tilde{\ell}}{\delta s}  + \frac{p^\prime}{F^2} \frac{\partial F}{\partial \eta } , \qquad    \frac{\delta \ell}{\delta p' } = 1 - \frac{\rho}{F} 
\end{equation}
recalling that $\eta = \frac{s}{\rho}$.

The equations of motion associated with this Lagrangian are
\begin{eqnarray}
\left( \partial_t + \pounds_u \right) \frac{\delta  \ell}{\delta u  } = \rho \nabla \frac{\delta  \ell}{\delta \rho  } + s \nabla \frac{\delta  \ell}{\delta s }\\
\bar D_t \rho =0, \qquad \bar D_t s=0
\end{eqnarray} 
plus the constraint equation
\begin{equation}
\label{constraint-eqn}
\frac{\delta  \ell}{\delta p'} = 1 - \frac{\rho}{F} = 0.
\end{equation}
Now use the functional derivative definitions to get
\begin{equation}\label{EP_tile_ell}
\left( \partial _t + \pounds_u \right) \frac{\delta \tilde{\ell}}{\delta u}  - \rho \nabla \frac{\delta \tilde{\ell}}{\delta \rho  }  - s \nabla \frac{\delta \tilde{\ell}}{\delta s  } + \rho \nabla \Big(\frac{p^\prime}{F} + \frac{p^\prime \eta}{F^2} \frac{\partial F}{\partial \eta } \Big) - s \nabla \Big(\frac{p^\prime}{F^2} \frac{\partial F}{\partial \eta } \Big)=0.
\end{equation}
Write the last two terms (after some algebra) as $\rho\, \mathcal{G} p^\prime$ with $\mathcal{G}$ the linear operator defined on functions $f$ as
\begin{equation}
\mathcal{G} f = \frac{1}{F} \nabla f - \frac{f}{F^2} \frac{\partial F}{\partial \phi }  \nabla \phi  = \nabla \Big(\frac{f}{F}\Big) + \frac{f}{F^2} \pp{F}{\eta} \nabla \eta.
\end{equation}
To determine the elliptic equation for $p^\prime$, we must discover the \textit{hidden constraints} associated with (\ref{constraint-eqn}), which are obtained by time differentiating twice. Start by writing (\ref{constraint-eqn}) as
\begin{equation}
\rho = F(\phi, \eta)
\end{equation}
and take $D_t$ of both sides to obtain
\begin{equation}
- \rho \operatorname{div}u  = (u\cdot \nabla \phi)\pp{F}{\phi}
\end{equation}
using $D_t \rho = - \rho \operatorname{div}u $, $D_t \phi = u \cdot \nabla \phi$ and $D_t \eta = 0$. When evaluated on the primary constraint $\rho = F(\phi, \eta)$ (and divided by $F^2$), this is just
\begin{equation}
\label{first-hidden}
- \frac{1}{F} \operatorname{div}u  = \frac{1}{F^2} (u \cdot \nabla \phi)\pp{F}{\phi}
\end{equation}
referred to as the \textit{first} hidden constraint. It can be written as
\begin{equation}
\label{first-hidden-constraint}
\mathcal{D} u = 0
\end{equation}
for the linear operator $\mathcal{D}$ defined by
\begin{equation}
\mathcal{D} v = \frac{1}{F} \operatorname{div}v  + \frac{1}{F^2} \pp{F}{\phi} (v \cdot \nabla \phi)= \frac{1}{F^2}  \operatorname{div} ( Fv) - \frac{1}{F^2} \pp{F}{\eta}(v \cdot \nabla \eta).
\end{equation}
Now time differentiate \eqref{first-hidden-constraint} to get
\begin{equation}
(\partial_t \mathcal{D}) u + \mathcal{D} \partial_tu = 0.
\end{equation}
The latter term can be expanded using 
\begin{equation}
\left( \partial_t + \pounds_ u \right) \Big(\frac{1}{\rho} \frac{\delta \tilde{\ell}}{\delta u  } \Big)  - \nabla \frac{\delta \tilde{\ell}}{\delta \rho  } - \eta \nabla \frac{\delta \tilde{\ell}}{\delta s } + \mathcal{G} p^\prime = 0,
\end{equation}
which directly follows from \eqref{EP_tile_ell}. If we note that $\frac{1}{\rho} \frac{\delta \tilde{\ell}}{\delta u}  = u+R$ for our Lagrangians, we have $ \partial _t\Big(\frac{1}{\rho} \frac{\delta \tilde{\ell}}{\delta u}\Big)= \partial _tu $, which gives finally 
\begin{equation}
\label{elliptic-eqn}
\mathcal{D} \mathcal{G} p^\prime = (\partial_t \mathcal{D}) u - \mathcal{D} \mathcal{E}
\end{equation}
with
\begin{equation}
\mathcal{E} = \pounds_ u  \Big(\frac{1}{\rho} \frac{\delta \tilde{\ell}}{\delta u  } \Big)  - \nabla \frac{\delta \tilde{\ell}}{\delta \rho  } - \eta \nabla \frac{\delta \tilde{\ell}}{\delta s }.
\end{equation}
This is the \textit{second} hidden constraint, and is the equation to determine $p^\prime$. Note that $\mathcal{D}$ and $\mathcal{G}$ are adjoints under a density-weighted inner product:
\begin{equation}
\int_\Omega \rho f \mathcal{D} v {\rm d}x + \int_\Omega \rho v \cdot \mathcal{G} f {\rm d} x = \int_{\partial \Omega} f v \cdot \hat{n}{\rm d}s,
\end{equation}
where we note that $\rho = F( \phi , \eta )$.

The operator $\mathcal{D} \mathcal{G}$ is \textit{elliptic}, which is easy to see since the 2nd order part is just
\begin{equation}
\frac{1}{F} \operatorname{div} \Big( \frac{1}{F} \nabla \square \Big)
\end{equation}
with $F > 0$. This is nothing more than the weighted Poisson operator, which is known to be elliptic. 

The boundary conditions are obtained by time differentiating the condition $u\cdot \hat{n} = 0$ on $\partial \Omega$. This ensures that if a velocity field initially satisfies $u \cdot \hat{n} = 0$ on $\partial \Omega$ and the constraint $\mathcal{D}u = 0$ on $\Omega$, then it does so for all time. From $ \partial _t u \cdot \hat{n} = 0$ this yields
\begin{equation}
\mathcal{G} p^\prime \cdot \hat{n} = - \mathcal{E}  \cdot \hat{n}
\end{equation}
on $\partial \Omega$.
With these boundaries conditions, the solution is unique up to $\delta p^\prime$ satisfying
\begin{equation}
    \mathcal{G} \delta p^\prime = 0
\end{equation}

To summarize, the elliptic pressure equation for a semi-compressible GFD fluid model with Lagrangian and constraints given in \eqref{general_ell_GFD} and \eqref{Constraint_C} is
\begin{equation}\label{Elliptlic_equation_reversible} 
\left\{
\begin{array}{l}
\vspace{0.2cm}\mathcal{D} \mathcal{G} p'= (\partial _t \mathcal{D}) u - \mathcal{D} \mathcal{E} \quad\text{on}\quad \Omega  \\
\mathcal{G} p^\prime \cdot \hat{n} = - \mathcal{E}  \cdot \hat{n} \quad\text{on}\quad \partial \Omega .
\end{array}
\right.
\end{equation}


\paragraph{Anelastic Equations.} Recalling that $F( \phi , \eta )= \rho  _0( \phi )$ in this case, the pressure equation \eqref{Elliptlic_equation_reversible} for the anelastic equations is obtained with
\begin{align*}
\mathcal{G} f &= \nabla \Big(\frac{f}{\rho_0(\phi)}\Big) \\
\mathcal{D}v &= \frac{1}{\rho_0(\phi)^2}\operatorname{div} \big(\rho_0(\phi) v\big) \\
\mathcal{E} &=  u \cdot \nabla u + 2 \omega \times u - b \nabla \phi.
\end{align*}

\paragraph{Pseudoincompressible Equations.} Recalling that $F( \phi , \eta )= \rho ^*(\phi, \eta ) = \rho  ( \bar p_0(\phi) , \eta )$ in this case, the pressure equation \eqref{Elliptlic_equation_reversible} for the pseudoincompressible is obtained with
\begin{align*}
\mathcal{G} f &=  \nabla \big(\frac{f}{\rho^*(\phi,\eta)}\big) - f \Gamma^*( \phi , \eta ) \nabla \eta = \frac{1}{\rho^*(\phi,\eta)} \Big(\nabla f  - f \kappa _{\rm ad}^*( \phi , \eta ) \nabla p_0\Big)\\
\mathcal{D} v&=\frac{1}{\rho^*(\phi, \eta)^2} \operatorname{div} \Big(\rho^*(\phi, \eta) v\Big) + \Gamma^*( \phi , \eta ) v\cdot \nabla \eta =\frac{1}{\rho^*(\phi, \eta)} \Big( \operatorname{div}v + \kappa _{\rm ad}^* ( \phi , \eta ) \nabla p_0 \cdot v  \Big) \\
\mathcal{E}&=  u \cdot \nabla u + 2 \omega \times u - b \nabla \phi ,
\end{align*}
where  $\Gamma^*(\phi, \eta) = \Gamma(\bar p_0(\phi), \eta) = \frac{\partial^2 h}{\partial p \partial \eta} (\bar p_0(\phi), \eta)$ and $ \kappa ^*_{\rm ad}( \phi , \eta )= \kappa _{\rm ad}( \bar p_0( \phi ), \eta ) = - \rho ^*( \phi , \eta ) \frac{\partial^2 h}{\partial p^2 } (\bar p_0(\phi), \eta)$. We note that $ \partial _t \mathcal{D} $ can be explicitly computed by using $\partial _t \eta = - v \cdot \nabla \eta $.

Now consider the special case of a perfect ideal gas. In this case $ \Gamma ^*= \frac{1}{ \rho  ^* C_p} $ and $ \kappa _{\rm ad}^* = \frac{1}{ p_0} \frac{C_v}{C_p}$ so the operator $ \mathcal{D} $ becomes
\[
\mathcal{D} v= \frac{1}{\rho^*(\phi, \eta)} \Big( \operatorname{div}v + \frac{1}{ p_0} \frac{C_v}{C_p}   \nabla  p_0 \cdot v  \Big)
\]
Furthermore, we have
\[
\rho  ^* \theta = \frac{1}{R} \left( \frac{1}{p_{00}} \right) ^{-R/C_p}\bar p_0( \phi )^{C_v/C_p}  
\]
from which we get
\[
\frac{C_v}{C_p} \frac{\nabla p_0}{p_0} = \frac{\nabla ( \rho  ^* \theta)}{ \rho  ^* \theta } 
\]
and hence
\[
\mathcal{D} v= \frac{1}{ \rho  ^*( \phi , \eta )}\Big(\operatorname{div}v+  \frac{\nabla ( \rho  ^* \theta) \cdot v}{ \rho  ^* \theta } \Big) = \frac{1}{ {\rho  ^*}^2 \theta } \operatorname{div}( \rho  ^* \theta v)  
\]
Now recall that for an ideal gas we have
\[
    p \alpha = \kappa  \theta \Pi \quad\quad \leftrightarrow \quad\quad \rho \theta = \frac{p}{\kappa \Pi} \quad\quad \leftrightarrow \quad\quad \rho(p, \eta) \theta(\eta) = \frac{p}{\kappa \Pi(p)},
\]
where $\Pi (p)=C_p (p/p_{00})^{R/C_p}$ is the Exner pressure and $ \kappa = R/C_p$. We prefer this definition of Exner pressure instead of the more common $\Pi (p)=(p/p_{00})^{R/C_p}$ since it naturally appears as the conjugate variable to potential temperature temperature $\theta$ when considering internal energy $U$; this property also generalizes to the case of moist air with condensates.

Evaluating both sides at $p =  p_0=\bar p_0( \phi )$ gives
\[
\rho^* \theta = \frac{ p_0}{\kappa \Pi_0}
\]
for $\Pi_0 = \Pi(\bar p_0( \phi ))$. Defining $ \theta _0$ through $p_0 \alpha _0= \kappa \Pi _0 \theta _0$, we have 
\[
\rho^* \theta = \rho_0 \theta_0
\]
and the reference state equation \eqref{rel_rho_0_p_0} reads $  \theta _0 \nabla \Pi _0 =- \nabla \phi $. We conclude that
\[
\mathcal{D} v = \frac{1}{ {\rho  ^*}^2 \theta } \operatorname{div}( \rho_0 \theta_0 v)  
\]
and therefore $\operatorname{div}( \rho_0 \theta_0 v) = 0$.

\subsubsection{Irreversible Dynamics}

Unfortunately, in the case of irreversible dynamics the derivation of the elliptic equation gets much more complicated. The principal reason for this is that $D_t \eta \neq 0$, and in fact it has terms involving $p^\prime$ in it through $j_s$ and $ \frac{\delta \ell}{\delta s} $. To work this out fully we need the expressions for $j_s$, etc, in terms of $ \frac{\delta \ell}{\delta s} $. The situation does significantly simplify if we have $F(\phi)$ instead of $F(\phi, \eta)$ since then $D_t \eta$ is not required and the previous approach basically follows through without issues, with only a slight modification of $\mathcal{E}$ required

To make this more explicit, consider the anelastic equations (\ref{BA_intermediate}), reproduced here for convenience:
\begin{equation}
\left\{
\begin{array}{l}
\vspace{0.2cm}\displaystyle
 \partial _t  u + u \cdot\nabla  u + 2 \omega \times u = - \nabla \frac{p'}{\rho_0} + b \nabla \phi  + \frac{1}{\rho}\operatorname{div}\sigma\\
\vspace{0.2cm}\displaystyle  T (\bar D_t s + \operatorname{div} j_s )  = \sigma  : \nabla u- j_s \cdot  \nabla T\\
\displaystyle \bar D_t \rho  =0, \qquad \rho  = \rho  _0.
\end{array}
\right.
\end{equation} 

The first equation (for $u$) in (\ref{BA_intermediate}) can be rewritten as 
\begin{equation}
\label{u-BA-irr}
\partial _t  u + u \cdot\nabla  u + 2 \omega \times u + \mathcal{G} p' - b \nabla \phi  - \frac{1}{\rho_0}\operatorname{div}\sigma=0
\end{equation}
for
\begin{eqnarray}
\mathcal{G} p' = \nabla \frac{p'}{\rho_0}.
\end{eqnarray}
Insertion of the primary constraint $\rho  = \rho  _0$ into the continuity equation $\bar D_t \rho  =0$ yields the first hidden constraint
\begin{equation}
\label{BA_first_hidden}
\mathcal{D} u = \frac{1}{ \rho  ^2_0} \operatorname{div} (\rho_0 u) = 0
\end{equation}
which is the same as in the reversible case. 

Since $\mathcal{D}$ is independent of time, time differentiation of (\ref{BA_first_hidden}) and substitution of (\ref{u-BA-irr}) yields 
\begin{eqnarray}
\operatorname{div} \Big(\rho_0  \nabla \frac{p'}{\rho_0}\Big) = - \operatorname{div} \Big(\rho_0 \Big( u \cdot\nabla  u + 2 \omega \times u - b \nabla \phi - \frac{1}{\rho_0}\operatorname{div}\sigma \Big)\Big)
\end{eqnarray}
which is exactly the same elliptic equation as found in the reversible case, with modified right hand to incorporate the $\frac{1}{\rho_0}\operatorname{div}\sigma$ term. No slip boundary condition can be treated via the pressure Poisson equation as in the usual case of the Navier-Stokes equations, see, e.g., \cite{ShRo2011} and references therein.

\section{Variational modelling of multicomponent semi-compressible fluids}
\label{sec-4}

Many real geophysical and astrophysical fluids are composed of multiple distinct components: moist air is a mixture of dry air, water vapor, liquid water and ice; seawater is a mixture of liquid water and salt, etc. Therefore, in this section we will review the variational formulation for fully compressible multicomponent fluids and then extend it to the case of semi-compressible multicomponent fluids. By component here we refer to thermodynamically distinct components within the fluid. This might include different phases of the same species (such as water vapor and liquid water) or even different \textit{allotropes} of the same species and phase (such as ortho and para forms of hydrogen). As in the previous sections, we will assume a closed manifold with material boundaries. Additionally, we will assume that the fluid is subject to \textit{local thermodynamic equilibrium} and that all components have the same temperature, and move with the same (barycentric) velocity. These are standard assumptions made in geophysical and astrophysical fluids. Although it is possible to relax them, this quickly becomes quite complicated, with separate velocity and entropy equations now required for each component.

\subsection{Variational formulation of multicomponent fluids}\label{subsec_VFMF}

\paragraph{Multicomponent fluids and their Lagrangians.} We consider a multicomponent fluid with $K$ components indexed by $k=1,...,K$.
Lagrangians of multicomponent fluids are functions of the form
\begin{equation}\label{Lagr_general_multi} 
L=L( \varphi , \dot \varphi , \{ \varrho _k\} ,S): T \operatorname{Diff}( \Omega ) \times \operatorname{Den}( \Omega ) ^K\times \operatorname{Den}( \Omega ) \rightarrow \mathbb{R},
\end{equation} 
with $\{ \varrho _k\}=(\varrho_1 ,..., \varrho _K)$ and where $\varrho_k   \in \operatorname{Den}( \Omega )$, $k=1,...,K$, denotes the mass density of the component $k$ in the material description. This Lagrangian satisfies the relabelling symmetry \eqref{Invariance_L} 
extended to several components and hence can be written in terms of Eulerian variables as in \eqref{L_ell} by
\[
L(\varphi , \dot{ \varphi }, \{ \varrho _k\} ,S)=\ell(u , \{ \rho _k\} ,s),
\]
with $\ell: \mathfrak{X} ( \Omega ) \times \operatorname{Den}( \Omega )^K \times \operatorname{Den}( \Omega ) \rightarrow \mathbb{R}$ the reduced (Eulerian) Lagrangian.

The fundamental GFD Lagrangian for multicomponent fluids takes the form
\begin{equation}\label{Fund_Lagr_multi} 
\ell(u,\{ \rho _k\}  , s) = \int_ \Omega \Big[ \frac{1}{2} \rho  | u| ^2 + \rho  R \cdot u - \rho  e( \rho  , \eta , \{ q _k\}) - \rho  \phi \Big]{\rm d} x,
\end{equation}
where $\rho  := \sum_k \rho  _k$ is the total mass density, $\eta := s/ \rho$ is the specific entropy, and $q_k:= \rho  _k/ \rho  $ is the concentration of component $k$, $k=1,...,K$.

\paragraph{Reversible case.} The variational formulation recalled in \S\ref{subsec_RFM} generalizes easily to multicomponent fluids, both in the Lagrangian and Eulerian formulations.

In the Lagrangian formulation, Hamilton's principle \eqref{HP} generalizes as
\begin{equation}\label{HP_multi} 
\left. \frac{d}{d\varepsilon}\right|_{\varepsilon=0}  \int_0^T L( \varphi _ \varepsilon  , \dot \varphi _ \varepsilon , \{\varrho _{k,0}\}, S_0){\rm d}t=0,
\end{equation}
where $ \varphi _ \varepsilon$ is an arbitrary path of time dependent diffeomorphisms prescribed at the temporal extremities $ t=0,T$. The mass densities $ \varrho _{k,0}$ and entropy density $S_0$ are held fixed and are time independent, as highlighted by the indices $0$.

In the Eulerian formulation, Hamilton's principle induces the Euler-Poincar\'e principle as a straightforward extension of \eqref{EP}--\eqref{EP_variation} to several components. The Euler-Poincar\'e system \eqref{system_Eulerian} becomes
\begin{equation}\label{system_Eulerian_multi} 
\left\{
\begin{array}{l}
\vspace{0.2cm}\displaystyle( \partial _t + \pounds _u) \frac{\delta \ell }{\delta u }= \sum_k\rho_k \nabla \frac{\delta \ell}{\delta \rho_k }+ s \nabla \frac{\delta\ell }{\delta s }\\
\bar D_t \rho_k =0, \;\;k=1,...,K,\qquad \bar D_t s=0.
\end{array}
\right.
\end{equation}

\paragraph{Irreversible case.} In addition to the irreversible processes considered in \S\ref{subsec_IFM},  we assume that the multicomponent fluid also involves diffusion described by the fluxes $J_k$ and intercomponent conversion described by the conversion rates $I_k$. In the case of a fluid containing multiple phases of the same species (for example, moist air), these intercomponent conversions would include the process of phase change.
Total mass conservation for these processes requires
\begin{equation}\label{mass_conservation_material} 
\sum_k J_k=0 \quad\text{and}\quad \sum_k I_k=0.
\end{equation} 
In addition to the internal entropy density variable $ \Sigma (t) \in \operatorname{Den}( \Omega )$ and the thermal displacement $ \Gamma (t) \in F( \Omega )$, the description of the additional processes also involves the matter displacements $W_k(t) \in  F( \Omega )$, $k=1,...,K$. Recall that in presence of irreversibility, we assume no-slip boundary conditions, i.e., $\varphi (t) \in \operatorname{Diff}_0( \Omega )$.

The variational formulation \eqref{VP_fluid}--\eqref{VC_fluid} extends naturally as \cite{GBYo2017b,GB2019}
\begin{equation}\label{VPLagr_atmosph} 
\delta   \int_0^T  \Big[L(\varphi, \dot\varphi, \{ \varrho _k\},S)  + \!\! \int_ \Omega \sum_k \varrho _k \dot W_k  {\rm d} X+  (S- \Sigma ) \dot \Gamma {\rm d} X\Big] \, {\rm d}t  =0,
\end{equation} 
subject to the \textit{phenomenological constraint}
\begin{equation}\label{KC_atmosphere}
\frac{\delta L }{\delta S}\dot \Sigma= - P: \nabla\dot \varphi  + J_S \cdot \nabla \dot\Gamma  + \sum_k(J _k \cdot \nabla \dot W _k +  I_k \dot W _k)
\end{equation}  
and with respect to variations $ \delta \varphi $, $ \delta S$, $ \delta \Sigma $, $ \delta \Gamma $ subject to the \textit{variational constraint}
\begin{equation}\label{VC_atmosphere}
\frac{\delta L}{\delta S}\delta \Sigma = - P: \nabla \delta \varphi  + J _S \cdot \nabla \delta\Gamma + \sum_k( J _k \cdot \nabla \delta  W _k+  I_k \delta  W _k)
\end{equation} 
and with $\delta \varphi$, $\delta\Gamma $, $\delta W_k$ vanishing at $t=0,T$.

A direct application of \eqref{VPLagr_atmosph}--\eqref{VC_atmosphere} yields the system
\begin{equation}\label{multicomp_L_material} 
\left\{
\begin{array}{l}
\vspace{0.2cm}\displaystyle\frac{d}{dt}\frac{\delta L}{\delta \dot \varphi }  - \frac{\delta L}{\delta \varphi } = \operatorname{DIV}P,\qquad \dot \varrho _k + \operatorname{DIV} J_k = I_k,\; \;k=1,...,K\\
\displaystyle-\frac{\delta L }{\delta S}(\dot S +\operatorname{DIV} J_S ) = P :\nabla\dot \varphi  + J_S \cdot \nabla \frac{\delta L }{\delta S}  + \sum_k \left(J _k \cdot \nabla \frac{\delta L }{\delta \varrho_k} + I_k \frac{\delta L }{\delta \varrho_k}\right),
\end{array}
\right.
\end{equation} 
together with the conditions
\begin{equation}\label{Additional_Conditions_Multi_Mat} 
\dot \Gamma = - \frac{\delta L }{\delta S}, \qquad \dot \Sigma =\dot S + \operatorname{DIV}J_S , \quad\text{and}\quad \dot W _k =- \frac{\delta L }{\delta \varrho_k }.
\end{equation}
The first two conditions are obtained as in the single component case from the variations $\delta S$ and $ \delta \Gamma $. In particular, $ \Gamma $ is the thermal displacement and $\dot  \Sigma $ is the rate of internal entropy production. The third condition imposes that $W_k$ is the thermodynamic displacement associated to transport of matter and follows from the variations $ \delta W_k$. If the variations $ \delta \Gamma $ and $ \delta W _k $ are free on $ \partial \Omega $, the variational formulation further imposes $J_s \cdot n=0$ and $J_k \cdot n=0$ on $ \partial \Omega $, i.e., the fluid domain is adiabatically closed. The second law of thermodynamics reads simply
\begin{equation}\label{2ndlaw_Lagrangian}
\dot\Sigma \geq 0
\end{equation}

The Eulerian versions of $W_k$, $J_k$, and $I_k$, are
\[
w_k= W_k \circ \varphi ^{-1} , \qquad j_k= ( \nabla \varphi \cdot J_k) \circ \varphi ^{-1} J \varphi ^{-1} , \qquad i_k = \varphi _* I_k
\] 
while the Eulerian version of the other variables have been described in \S\ref{subsec_IFM}. Total mass conservation requires
\begin{equation}\label{mass_conservation_Eulerian} 
\sum_k j_k=0 \quad\text{and}\quad \sum_ki_k=0,
\end{equation}
which follows from \eqref{mass_conservation_material}. 
From this, the Eulerian version of the variational formulation \eqref{VPLagr_atmosph}--\eqref{VC_atmosphere} becomes
\begin{equation}\label{Review_GBYo} 
\delta   \int_0^T \Big[\ell(u,\{ \rho _k\}, s)+\int_\Omega \sum_k \rho _k D_t w_k   {\rm d}x + (s- \varsigma  ) D_t \gamma    {\rm d}x \Big] {\rm d}t =0,
\end{equation}
subject to the \textit{phenomenological constraint}
\begin{equation}\label{KC_GBYo}
\frac{\delta \ell}{\delta s}\bar D_t \varsigma =  - \sigma : \nabla u  + j _s \cdot \nabla D_t\gamma  + \sum_k( j _k \cdot \nabla D_tw _k+  i_k D_t w _k )
\end{equation}
and with respect to variations subject to $ \delta u = \partial _t \zeta+[ u , \zeta ] $ and to the \textit{variational constraint}
\begin{equation}\label{VC_GBYo}
\frac{\delta \ell}{\delta s}\bar D_\delta \varsigma = - \sigma : \nabla \zeta + j _s \cdot \nabla D_\delta\gamma  + \sum_k(j _k \cdot \nabla D_\delta w _k+  i_kD_\delta w _k )
\end{equation} 
with $ \delta w _k $, $ \delta \gamma $, and $ \zeta  $ vanishing at $t=0,T$.

A direct application of \eqref{Review_GBYo}--\eqref{VC_GBYo} yields the system
\begin{equation}\label{system_Eulerian_atmosphere_moist_L} 
\left\{
\begin{array}{l}
\vspace{0.2cm}\displaystyle
\partial _t \frac{\delta \ell}{\delta u}  + \pounds _ u \frac{\delta \ell}{\delta u} = \sum_k\rho _k \nabla \frac{\delta \ell}{\delta \rho _k }+ s \nabla \frac{\delta \ell }{\delta s }+\operatorname{div}  \sigma \\
\vspace{0.2cm}\displaystyle - \frac{\delta \ell}{\delta s } (\bar D_t s + \operatorname{div} j _s ) =  \sigma \!: \!\nabla u + j_s\! \cdot  \!\nabla \frac{\delta \ell }{\partial s } + \sum_k\Big(j _k \!\cdot \! \nabla \frac{\delta \ell}{\delta \rho _k }+i_k \frac{\delta \ell }{\delta \rho_k} \Big) \\
\displaystyle \bar D_ t\rho _k+  \operatorname{div}  j_k=i _k,\;\;k=1,...,K,
\end{array}
\right.
\end{equation}
with boundary conditions $u|_{ \partial \Omega }=0$. Exactly as in the material description in \eqref{Additional_Conditions_Multi_Mat}, the principle also yields the conditions
\begin{equation}\label{additional_conditions_multi} 
D_t \gamma = - \frac{\delta  \ell }{\delta  s}, \qquad \bar D_t \varsigma = \bar D_t s + \operatorname{div} j_s, \quad\text{and}\quad  D_tw_k= -\frac{\delta \ell }{\delta \rho_k} \quad\text{on}\quad \Omega 
\end{equation}
arising from the variations $ \delta s$, $ \delta \gamma$, and $ \delta \varrho _k$. If the variation $ \delta \gamma $ and $ \delta w _k $ are free on $ \partial \Omega $, the variational formulation imposes
\[
j_k \cdot n =0 \qquad\text{and}\qquad j_s \cdot n =0 \quad \text{on} \quad  \partial \Omega,
\]
i.e., the fluid domain is adiabatically closed. We refer to \cite{GBYo2017b,GB2019} for detailed computations. The second law of thermodynamics in an Eulerian description reads simply
\begin{equation}\label{2ndlaw_Eulerian}
\bar D_t \varsigma  \geq 0.
\end{equation}
System \eqref{system_Eulerian_atmosphere_moist_L} gives the general equations of motion for a fluid with Lagrangian $\ell(u,\{\rho_k\},s)$ subject to the irreversible processes of viscosity, heat conduction, diffusion, and phase changes.

The system is closed by specifying a relationship, or \textit{parameterizing}, the thermodynamic fluxes ($\sigma $, $j_s$, $j_k$, $i_k$) in terms of the thermodynamic forces ($\operatorname{Def}u= \frac{1}{2}(\nabla u+\nabla u^\mathsf{T})$, $\nabla T$, $\nabla \mu_k$, $\mu_k$), see \cite{GB2019}.

For the Lagrangian \eqref{Fund_Lagr_multi} of the rotating compressible multicomponent fluid, \eqref{system_Eulerian_atmosphere_moist_L} yields
\begin{equation}\label{equa_multi_comp_fluid} 
\left\{
\begin{array}{l}
\vspace{0.2cm}\displaystyle
\rho(\partial _t u+ u\cdot\nabla u + 2 \omega \times u) = - \nabla p - \rho \nabla \phi+  \operatorname{div} \sigma \\
\displaystyle T (\bar D_t s +  \operatorname{div} j _s ) =   \sigma :  \nabla u - j _s \cdot \nabla T -  \sum_k\left(j _k\cdot \nabla \mu _k + i _k  \mu_k\right) \\
\displaystyle \bar D_ t\rho _k+  \operatorname{div}j _k=i_k,
\end{array}
\right.
\end{equation}
where use of \eqref{mass_conservation_Eulerian} was made. 

\paragraph{Enthalpy and Gibbs potential.} Equations \eqref{equa_multi_comp_fluid} can also be obtained by using the pressure or, alternatively, the pressure and temperature as independent variables. Exactly as in \S\ref{h_and_g} for the single component case, this is achieved by expressing the Lagrangian in terms of the enthalpy $h(p, \eta ,\{ q_k\})$ or the Gibbs potential $g(p,T,\{ q_k\})$ for multicomponent fluids.
Expression \eqref{ell_enth} and \eqref{ell_gibbs} become
\begin{equation}\label{ell_enth_multi} 
\ell_h(u, \{ \rho _k\}  , s,p ) = \int_ \Omega \Big[ \frac{1}{2} \rho  | u| ^2 + \rho  R \cdot u - \rho  h( p  , \eta , \{ q _k\}) +p - \rho  \phi \Big]{\rm d} x
\end{equation} 
and
\begin{equation}\label{ell_gibbs_multi} 
\ell_g(u,\{ \rho _k\}, s,p,T ) = \int_ \Omega \Big[ \frac{1}{2} \rho  | u| ^2 + \rho  R \cdot u - \rho  g( p  , T , \{q_k\}) +p - T s  - \rho  \phi \Big]{\rm d} x,
\end{equation}
where as above $\rho  := \sum_k \rho  _k$, $\eta := s/ \rho$, and $q_k:= \rho  _k/ \rho  $.
The variational formulation \eqref{Review_GBYo}--\eqref{VC_GBYo} generalizes to such Lagrangians by including the criticality condition with respect to arbitrary variations of the additional variables $p$ or, alternatively, $p$ and $T$.

\begin{remark}[Intrinsic formulation of multicomponent fluids on manifolds]\label{Remark_Manifolds_multicomponent}{\rm As mentioned in Remark \ref{Remark_Manifolds}, all the results of this paper can be written for fluids on general Riemannian manifold. Similarly with the single component case in \eqref{NSF_GFD_Eulerian_manifolds}, system \eqref{equa_multi_comp_fluid} can be written on Riemannian manifolds as
\begin{equation}\label{NSF_GFD_Eulerian_manifolds_multi} 
\left\{
\begin{array}{l}
\vspace{0.2cm}\displaystyle
\rho \big(  \partial_t u + u \cdot \nabla u +  2( i _ u \omega   )^\sharp\big) = - \operatorname{grad} p - \rho \operatorname{grad}\phi +\operatorname{div} \sigma \\
\displaystyle  T (\bar D_t s + \operatorname{div}j _s) = \sigma  ^\flat : \nabla u - j _s \cdot  {\rm d}  T-\sum_k (j_k \cdot {\rm d} \mu _k + i_k \mu _k )\\
\displaystyle \bar D_t \rho_k + \operatorname{div}j_k =i_k,
\end{array}
\right.
\end{equation}
with the same definition of the operators $ \nabla $, $ \operatorname{div}$, $ \operatorname{grad}$, and $\bar D_t$ in terms of the Riemannian metric as in in Remark \ref{Remark_Manifolds}, while ${\rm d}$ is the differential.}
\end{remark}

\subsection{Multicomponent pseudoincompressible model}

\paragraph{Lagrangian for multicomponent pseudoincompressible models.} This is obtained exactly as in the single component case by linearizing the Lagrangian $\ell_h$ in \eqref{ell_enth_multi} or $\ell_g$ in \eqref{ell_gibbs_multi}  around the background pressure  $p_0(x)$. One gets the Lagrangians
\begin{equation}\label{ell_PI_multi}
\begin{aligned} 
&\ell_{\rm pi}(u, \{ \rho _k\} , s, p')\\
&= \int_ \Omega \Big[ \frac{1}{2} \rho  | u| ^2 + \rho  R \cdot u  - \rho  h(p_0, \eta , \{ q _k\})- \rho  \phi  +p' \Big( 1 - \frac{\rho  }{ \rho  (p_0, \eta,\{ q _k\})} \Big) + p_0 \Big]{\rm d} x
\end{aligned} 
\end{equation} 
and 
\begin{equation}\label{ell_PI_g_multi}
\begin{aligned}
&\ell_{\rm pi}(u, \{ \rho _k\}  , s, p', T)\\
&=\int_ \Omega \Big[ \frac{1}{2} \rho  | u| ^2 + \rho  R \cdot u  - \rho  g(p_0,T,\{q _k\} )- \rho  \phi  -sT +p' \Big( 1 - \frac{\rho  }{ \rho  (p_0, T ,\{ q _k\})} \Big) +p_0 \Big]{\rm d} x.
\end{aligned} 
\end{equation}
In \eqref{ell_PI_multi} the expression $ \rho  (p_0, \eta ,\{ q _k\})$ is obtained as  $\frac{1}{\rho  (p_0, \eta,\{ q _k\}) }= \frac{\partial h}{\partial p}(p_0, \eta , \{q _k\})$ and in \eqref{ell_PI_g_multi} as $\frac{1}{ \rho  (p_0,T, \{ q _k\})}= \frac{\partial g}{\partial p}(p_0, T, \{q _k\})$.

\paragraph{Variational derivation of multicomponent pseudoincompressible thermodynamics.} We apply the variational formulation \eqref{Review_GBYo}--\eqref{VC_GBYo}  to the Lagrangian \eqref{ell_PI_multi} with arbitrary variations $ \delta p'$. This gives the system \eqref{system_Eulerian_atmosphere_moist_L} with $\ell$ replaced by $\ell_{\rm pi}$ and with the condition
\[
\frac{\delta \ell_{\rm pi}}{\delta p'}=0 .
\] 
The derivatives of the Lagrangian \eqref{ell_PI_multi} are computed as
\begin{equation}\label{functional_der_pi} 
\begin{aligned} 
&\frac{\delta \ell_{\rm pi}}{\delta u}= \rho  (u+R), \qquad \frac{\delta \ell_{\rm pi}}{\delta s} =  - \frac{\partial h}{\partial \eta } (p_0, \eta, \{q_k\} )-  p'  \Gamma (p_0, \eta, \{q_k\} )\\
&\frac{\delta \ell_{\rm pi}}{\delta p'}= 1 - \frac{\rho  }{ \rho  (p_0, \eta , \{q_k\})}\\
&\frac{\delta \ell_{\rm pi}}{\delta \rho_k}  = \frac{1}{2} | u| ^2 + R \cdot u - \frac{\partial h}{\partial q_k}(p_0, \eta , \{q_k\}) - \phi -  p' \frac{\partial ^2 h}{\partial p \partial q _k }(p_0, \eta ,\{q_k\}),
\end{aligned}
\end{equation}
$k=1,...,K$, where $ \Gamma (p, \eta,\{q_k\} )= \frac{\partial ^2 h}{\partial p \partial \eta }(p, \eta, \{q_k\} )$ is the adiabatic temperature gradient. As in the single component case, the \textit{modified temperature} naturally appears as
\[
T^*:= - \frac{\delta \ell_{\rm pi}}{\delta s} =  T(p_0, \eta, \{q_k\} ) + p'  \Gamma (p_0, \eta, \{q_k\} ).
\]
The novelty here is the occurrence of the \textit{modified chemical potential} which follows from the expression of the derivative $ \frac{\delta \ell}{\delta \rho  _k }$, namely
\[
\mu _k ^* =  \frac{\partial h}{\partial q_k}(p_0, \eta , \{q_k\}) + p' \frac{\partial ^2 h}{\partial p \partial q _k }(p_0, \eta ,\{q_k\})= \mu _k + p' \frac{\partial ^2 h}{\partial p \partial q _k }(p_0, \eta ,\{q_k\}).
\]
The occurrence of these modified thermodynamic fluxes in the entropy equation, rather than the usual ones, follows from the first and third conditions in \eqref{additional_conditions_multi}, which arise from the variations $ \delta s$ and $ \delta \rho  _k $, 

We assume as earlier
\begin{equation}\label{mass_conservation_pi} 
\sum_k j_k=0 \quad\text{and}\quad \sum_k i _k =0
\end{equation} 
hence the total mass $ \rho  =\sum_k \rho  _k$ satisfies
\[
\bar D_t \rho  =0.
\]
With this assumption, the variational formulation yields the system of equations
\begin{equation}\label{Eulerian_PI_multi} 
\left\{
\begin{array}{l}
\vspace{0.2cm}\displaystyle
\rho  ( \partial _t u + u \cdot \nabla u + 2 \omega  \times u) = - ( \rho  - \rho  _0) \nabla \phi - \nabla p' + p' \kappa _{\rm ad}\nabla p_0 +\operatorname{div}  \sigma \\
\vspace{0.2cm}\displaystyle \rho  = \rho  (p_0, \eta , \{q_k\}) \\
\displaystyle T^*(\bar D_ts  + \operatorname{div} j _s) =  \sigma : \nabla u - j _s\cdot  \nabla T^* - \sum_k\left(j _k\cdot \nabla \mu _k ^*+ i _k  \mu_k^*\right) \\
\displaystyle \bar D_ t\rho _k+  \operatorname{div}j _k=i_k,\;\; k=1,...,K,
\end{array}
\right.
\end{equation}
where $ \kappa _{\rm ad}= \frac{1}{ \rho  c_s ^2 }= - \rho  \frac{\partial ^2 h}{\partial p ^2 } (p, \eta , \{q_k\})$ is the adiabatic compressibility coefficient, evaluated at $(p_0(x), \eta , \{q_k\})$. This derivation also uses the relation \eqref{rel_rho_0_p_0}.
System \eqref{Eulerian_PI_multi} form a closed system for $u, \eta , \rho  _1,..., \rho  _K, p'$. Some steps of this derivation are presented in Appendix \ref{appendix_A}.


\paragraph{Energy conservation.} The total energy density of the multicomponent pseudoincompressible model is obtained as in the single component case as
\[
e_{\rm pi}= \frac{1}{2} \rho  | u| ^2 + \rho  h(p_0, \eta, \{q_k\} ) - p _0 + \rho  \phi  - p' \left( 1 - \frac{\rho  }{ \rho  (p_0, \eta,\{q_k\} )} \right),
\]
with the last term vanishing since that is just the soundproofing constraint.
A long, but straightforward computation yields the energy conservation equation
\[
\bar D_t e_{\rm pi}= \operatorname{div} \Big(  \sigma \cdot u  - p u - j_s T^* -\sum_k j_k \mu _k ^* \Big)
\]
along the solutions of \eqref{Eulerian_PI_multi}, with $p=p_0+p'$ and where one notes the occurrence of the modified temperature and modified chemical potentials $T^*$ and $ \mu _k ^* $. The total energy \eqref{E_tot} thus satisfies
\[
\frac{d}{dt} E_{\rm pi}= 0
\]
since $u |_ { \partial \Omega }=0$ and if $j_s \cdot n=0$ and $j_k \cdot n=0$, i.e., the fluid does not exchange work, heat, and matter with the exterior, consistently with the first law of thermodynamics.

\paragraph{Entropy production and phenomenological relations.} The system of equations \eqref{Eulerian_PI_multi} needs to be supplemented with phenomenological expressions for the thermodynamic fluxes in terms of the thermodynamic forces compatible with the second law \eqref{2ndlaw_Eulerian} which, for multicomponent pseudoincompressible fluids, takes the form
\[
\bar D_t \varsigma =I =\frac{1}{T^*} \Big( \sigma  : \nabla v  - j _s  \cdot \nabla  T^*  - \sum_k j _k  \cdot \nabla \mu _k ^* - \sum_k i _k \mu _k^* \Big)\geq 0.
\]
This form of entropy production is well-adapted for an application of the Onsager relations. Decomposing $ \sigma $ and $ \operatorname{Def}u$ into traceless and diagonal parts $ \sigma = \sigma ^{(0)} + \frac{1}{3} \operatorname{Tr}( \sigma ) \delta  $ and $ \operatorname{Def}u = ( \operatorname{Def}u)^{(0)} + \frac{1}{3} (\operatorname{div}u) \delta$, the linear phenomenological relations take the usual form in which $T$ and $ \mu _k$ are replaced by their modified expressions $T^*$ and $ \mu _k^*$:
\begin{align*} 
\vspace{0.2cm}-\begin{bmatrix}
\vspace{0.05cm}j_s\\
\vspace{0.05cm} j_k
\end{bmatrix}=\,&\, 
\begin{bmatrix}
L _{ss} & L _{sk}  \\[0.2em]
L _{ks} & L _{kk}  \\
\end{bmatrix}
\begin{bmatrix}
\vspace{0.05cm} \nabla  T^*\\
\vspace{0.05cm} \nabla \mu^*_k
\end{bmatrix}\\
\vspace{0.2cm}\begin{bmatrix}
\operatorname{Tr} \sigma  \\[0.2em]
- i_k
\end{bmatrix}= \,&\,
\begin{bmatrix}
\mathcal{L} _{00} & \mathcal{L} _{0k} \\[0.2em]
\mathcal{L} _{k0}&\mathcal{L} _{kk}
\end{bmatrix}
\begin{bmatrix}
\vspace{0.05cm}\frac{1}{3} \operatorname{div} {\bf v}  \\
\vspace{0.05cm}  \mu ^* _k
\end{bmatrix}\\
\sigma^{(0)} =\,&\,2 \mu (\operatorname{Def}u)^{(0)} 
\end{align*} 
where the matrices $L$ and $\mathcal{L}$ are positive, satisfy the Onsager-Casimir relations, as well as \eqref{mass_conservation_pi}, see \cite{dGMa1969}. Following \cite{ElGB2018}, it would also be possible to also treat the irreversible thermodynamic fluxes as a particular type of turbulence model, and obtain a set of turbulent fluxes that conform to a version of the 1st and 2nd laws of thermodynamics.

\paragraph{Buoyancy formulation.} For the multicomponent case, the buoyancy is defined is in \eqref{def_buoyancy} by
\begin{equation}\label{def_buoyancy_multi} 
b( \phi  , \eta , \{q_k\}):= - \frac{\partial }{\partial \phi } \big( h(\bar p_0( \phi  ), \eta,  \{q_k\} )+ \phi  \big) = \frac{ \bar\rho  _0( \phi  ) -  \rho  (\bar p_0( \phi ), \eta,  \{q_k\} )}{ \rho  (\bar p_0( \phi ), \eta,  \{q_k\} ) }.
\end{equation} 
Relation \eqref{differential_b} is 
\begin{equation}\label{differential_b_multi} 
{\rm d} b =  \frac{1}{\rho}\big(\rho  _0 ^2\kappa _{\rm ad}   +  {\bar\rho  _0}'( \phi )\big) {\rm d} \phi  +  \rho  _0  \Gamma {\rm d} \eta +\sum_k \rho  _0 \frac{\partial ^2 h}{\partial p \partial q _k  } {\rm d} q _k ,
\end{equation}
where $ \rho  $, $ \kappa _{\rm ad}$, and $ \Gamma $ are all expressed at $(\bar p_0( \phi ), \eta , \{q_k\})$.  From the entropy equation in \eqref{Eulerian_PI_multi} and \eqref{def_buoyancy_multi}, we get  the buoyancy equation as
\begin{equation}\label{buoyancy_equ_PI_multi} 
\begin{aligned} 
D_t b &=\frac{1}{\rho}\big( \rho  _0 ^2\kappa _{\rm ad}   +  {\bar\rho  _0}'( \phi )\big) D_t\phi  + \frac{\rho  _0 \Gamma }{\rho  T^*} \Big(   \sigma : \nabla u - \operatorname{div} ( j _s   T^*) - \sum_k(j_k \cdot \nabla \mu _k ^* + i _k \mu _k ^*  )\Big) \\
& \quad +\sum_k \frac{\rho  _0}{ \rho  } \frac{\partial ^2 h}{\partial p \partial q _k  }(i_k - \operatorname{div}j_k).
\end{aligned} 
\end{equation} 
System \eqref{Eulerian_PI_multi} can be expressed for the buoyancy $b$ as a prognostic variable instead of the specific entropy $ \eta $ by replacing the entropy equation of \eqref{Eulerian_PI_multi} with the buoyancy equation \eqref{buoyancy_equ_PI} and using the relation $ \eta = \eta ( \phi ,b, \{q _k\})$ to rewrite the expressions $ \Gamma (\bar p_0( \phi ), \eta , \{q _k\})$, $ \kappa _{\rm ad}(\bar p_0( \phi ), \eta, \{q _k\} )$, $T^*( \bar p_0( \phi ), \eta , \{q _k\})$, $ \mu _k ^* ( \bar p_0( \phi ), \eta , \{q _k\})$ and $ \rho  (\bar p_0( \phi ), \eta, \{q _k\} )$ in terms of $(\phi ,b, \{q _k\})$. In particular, we have
\[
\rho( \phi ,b, \{q _k\}) = \bar{\rho } _0( \phi ) \frac{1}{b+ 1}
\]
and the momentum equation in \eqref{Eulerian_PI_multi} takes the form
\[
\partial _t u + u \cdot \nabla u + 2 \omega  \times u = b \nabla \phi - \frac{1}{ \rho  } \nabla p' + \frac{\kappa _{\rm ad}}{ \rho  }p' \nabla p_0 + \frac{1}{ \rho  }\operatorname{div}  \sigma.
\]

\paragraph{Temperature and potential temperature formulations.} To obtain the equation in terms of the temperature, we consider $T$ as a function $T( \phi , \eta , \{q_k\})= T(\bar p_0( \phi ), \eta, \{ q_k\} )$ and the differential relation
\begin{equation}\label{differential_T_multi} 
{\rm d}T = - \rho  _0 \Gamma {\rm d} \phi  + \frac{T}{C_p}{\rm d} \eta + \sum_k \frac{\partial ^2 h}{\partial \eta \partial q _k }{\rm d} q _k  ,
\end{equation} 
where $ \Gamma $ and $C_p$ are evaluated at $(\bar p_0( \phi ), \eta, \{q_k\} )$.
This gives the temperature equation
\begin{equation}\label{temperature_equ_PI_multi}
\begin{aligned} 
D_t T &=  -  \rho  _0 \Gamma D_t \phi + \frac{T}{ \rho  \, C_p T^*} \Big(  
\sigma : \nabla u - \operatorname{div} ( j _s  T^*)- \sum_k(j_k \cdot \nabla \mu _k ^* + i _k \mu _k ^*  )\Big) \\
& \quad +\sum_k \frac{1}{ \rho  } \frac{\partial ^2 h}{\partial p \partial q _k  }(i_k - \operatorname{div}j_k).
\end{aligned} 
\end{equation}

\medskip 

To obtain the equation in terms of the potential temperature defined by $ \theta ( \eta ,\{q_k\})= T(p_{00}, \eta, \{q_k\} )$ for some reference value $p_{00}$ of the pressure, we use the differential relation
\begin{equation}\label{differential_theta_multi} 
{\rm d} \theta = \frac{\theta}{C_p(p_{00}, \eta , \{q_k\} ) }{\rm d} \eta  + \sum_k \frac{\partial ^2 h}{\partial \eta \partial q _k  }(p_{00},  \eta , \{q_k\}) {\rm d} q _k 
\end{equation} 
which gives
\[
\begin{aligned}
D_ t\theta &= \frac{ \theta }{ \rho  C_p(p_{00}, \theta  ,\{q_k\})  T^*} \Big(  
\sigma : \nabla u - \operatorname{div} ( j _s  T^*)- \sum_k(j_k \cdot \nabla \mu _k ^* + i _k \mu _k ^*  )\Big) \\
& \quad +\sum_k \frac{1}{ \rho  }\frac{\partial ^2 h}{\partial \eta  \partial q _k  }(p_{00}, \eta , \{q_k\})(i_k - \operatorname{div}j_k)
\end{aligned} 
\]
and allows to express system \eqref{Eulerian_PI_multi} in terms of $ \theta $ instead of $ \eta $.

\paragraph{General entropic variable $\chi$.} As in the single component case, we finish by considering the case of a general entropic variable $\chi = \chi(\eta, q_k)$ that is a function of specific entropy and mixing ratios. Noting the differential relationship
\begin{equation}
{\rm d} \chi = \pp{\chi}{\eta}{\rm d} \eta  + \sum_k \pp{\chi}{q_k} {\rm d} q _k 
\end{equation}
we obtain
\[
\begin{aligned}
D_ t\chi &= \pp{\chi}{\eta} \frac{ 1 }{\rho   T^*} \Big(  
\sigma : \nabla u - \operatorname{div} ( j _s  T^*)- \sum_k(j_k \cdot \nabla \mu _k ^* + i _k \mu _k ^*  )\Big) \\
& \quad +\sum_k \frac{1}{ \rho  } \pp{\chi}{q_k} (i_k - \operatorname{div}j_k)
\end{aligned} 
\]
As before, potential temperature $\theta$ is in fact one example of this.

\subsection{Multicomponent anelastic and Boussinesq models}

\paragraph{Lagrangian for multicomponent anelastic models.} This is obtained by replacing the condition $ \rho  = \rho  (p_0, \eta , \{q_k\})$ by $ \rho  = \rho  _0$ in the pseudoincompressible constraint. With this the Lagrangian \eqref{ell_PI_multi} becomes
\begin{equation}\label{ell_an_multi}
\begin{aligned} 
\ell_{\rm an}(u, \{ \rho _k\} , s, p') = \int_ \Omega \Big[ \frac{1}{2} \rho  | u| ^2 + \rho  R \cdot u  - \rho  h(p_0, \eta , \{ q _k\})- \rho  \phi  +p' \Big( 1 - \frac{\rho  }{ \rho_0} \Big) +p_0 \Big]{\rm d} x
\end{aligned} 
\end{equation} 
similarly for \eqref{ell_PI_g_multi} when using the Gibbs description.

\paragraph{Variational derivation of multicomponent anelastic thermodynamics.} We apply the variational formulation \eqref{Review_GBYo}--\eqref{VC_GBYo}  to the Lagrangian \eqref{ell_an_multi} with arbitrary variations $ \delta p'$. This yields
\begin{equation}\label{BA_final_multi} 
\left\{
\begin{array}{l}
\vspace{0.2cm}\displaystyle
 \partial _t  u + u \cdot\nabla u + 2 \omega \times u = b\nabla \phi  - \nabla \frac{p'}{\rho_0} + \frac{1}{\rho_0}\operatorname{div}  \sigma \\
\displaystyle T(\bar D_ts  + \operatorname{div} j _s) =  \sigma : \nabla u - j _s\cdot  \nabla T - \sum_k\left(j _k\cdot \nabla \mu _k + i _k  \mu_k\right) \\
\vspace{0.2cm}\displaystyle \bar D_ t\rho _k+  \operatorname{div}j _k=i_k,\;\; k=1,...,K,\\
\displaystyle \rho  = \rho  _0,
\end{array}
\right.
\end{equation}
where the buoyancy $b$ is given in \eqref{def_buoyancy}. From $ \rho  = \sum_k \rho  _k $ and $\bar D_t \rho  =0$, we get $\operatorname{div}( \rho_0  u)=0$.

The entropy equation can be rewritten in terms of $b$, $T$, $ \theta $ or $\chi$ as
\[
\begin{aligned} 
D_t b &=\frac{1}{\rho}\big( \rho  _0 ^2\kappa _{\rm ad}   +  {\bar\rho  _0}'( \phi )\big) D_t \phi  + \frac{\Gamma }{ T} \Big(   \sigma : \nabla u - \operatorname{div} ( j _s   T) - \sum_k(j_k \cdot \nabla \mu _k + i _k \mu _k  )\Big) \\
& \quad +\sum_k \frac{\partial ^2 h}{\partial p \partial q _k  }(i_k - \operatorname{div}j_k),
\end{aligned} 
\]
\[
\begin{aligned} 
D_t T &=  - \rho  _0 \Gamma D_t \phi + \frac{1}{ \rho _0 C_p } \Big(  
\sigma : \nabla u - \operatorname{div} ( j _s  T)- \sum_k(j_k \cdot \nabla \mu _k  + i _k \mu _k  )\Big) \\
& \quad +\sum_k \frac{1}{ \rho_0  } \frac{\partial ^2 h}{\partial p \partial q _k  }(i_k - \operatorname{div}j_k),
\end{aligned} 
\]
\[
\begin{aligned}
D_ t\theta &= \frac{ \theta }{ \rho_0  C_p(p_{00}, \theta  ,\{q_k\})  T} \Big(  
\sigma : \nabla u - \operatorname{div} ( j _s  T)- \sum_k(j_k \cdot \nabla \mu _k  + i _k \mu _k)\Big) \\
& \quad +\sum_k \frac{1}{ \rho_0}\frac{\partial ^2 h}{\partial \eta  \partial q _k  }(p_{00}, \eta , \{q_k\})(i_k - \operatorname{div}j_k),
\end{aligned} 
\]
\[
\begin{aligned}
D_ t\chi &= \pp{\chi}{\eta} \frac{1}{ T} \Big(  
\sigma : \nabla u - \operatorname{div} ( j _s  T)- \sum_k(j_k \cdot \nabla \mu _k  + i _k \mu _k)\Big) \\
& \quad +\sum_k \frac{1}{ \rho_0} \pp{\chi}{q_k} (i_k - \operatorname{div}j_k),
\end{aligned} 
\]
where all functions are expressed in terms of $(p_0, b, \{q_k\})$, $(p_0, T, \{q_k\})$, or, $(p_0, \theta , \{q_k\})$.

\subsection{Inclusion of chemical reactions}
For both geophysical and astrophysical applications, an additional set of important irreversible processes are chemical reactions. These can be naturally introduced in the semi-compressible fluid models discussed in this paper, by following the variational formulation for multicomponent reacting fluids in \cite{GBYo2017b}.

We assume that the multicomponent fluid undergoes $a=1,...,r$ chemical reactions, denoted 
\[
\sum_k\nu'_{ak}\,k\; \stackrel[a _{(2)}]{ a _{(1)} }{\rightleftarrows} \; \sum_k\nu ''_{ak}\, k, \quad a=1,...,r,
\]
with $a_{(1)}$, $a_{(2)}$ the forward and backward reactions and 
$\nu'_{ak}$, $\nu''_{ak}$ the forward and backward stoichiometric coefficients for the species $k$ in reaction $a$. Mass conservation during each reaction arises from the condition
\begin{equation}\label{Lavoisier_law}
\sum_k \bar \nu _{ak}=0, \qquad a=1,..., r \quad \text{(Lavoisier law)},
\end{equation} 
where $\bar\nu _{ak}= \nu _{ak}m_k$ with $m_k$ the molar mass of component $k$.
The thermodynamic force and thermodynamic flux for a chemical reaction are the chemical affinity $ \mathcal{A}_a = - \sum_k \bar\nu _{ak}  \mu _k$\footnote{The occurrence of $\bar\nu_{ak}= \nu _{ak}m_k$, rather than $\nu_{ak}$, in the definition of the chemical affinity arises since the chemical potential $ \mu _k$ is given per unit of mass rather than mole, see \cite{GBYo2017b}. We are using here mass densities $ \rho  _k= m_kn_k$ rather than molar densities $n_k$ like in \cite{GBYo2017b}.} and the reaction rate $j_a$, $a=1,...,r$.

The variational formulation \eqref{Review_GBYo}--\eqref{VC_GBYo} is extended to reacting fluids as follows, where, for simplicity, we ignore conversion rates $i_k$ due to other processes than chemical reactions. In addition to the thermodynamic displacements $\gamma $ and $ w_k$, we need to consider the thermodynamic displacements $\upsilon_a$ associated to chemical reaction $a$. For the pseudoincompressible case with chemical reactions, taking the Lagrangian $\ell_{\rm pi}$ in \eqref{ell_PI_multi}, we get the following variational formulation:
\begin{equation}\label{Review_GBYo_reaction} 
\delta   \int_0^T \Big[\ell_{\rm pi}(u,\{ \rho _k\}, s)+\int_\Omega \sum_k \rho _k D_t w_k   {\rm d}x + (s- \varsigma  ) D_t \gamma    {\rm d}x \Big] {\rm d}t =0,
\end{equation}
subject to the \textit{phenomenological and chemical constraints}
\begin{equation}\label{KC_GBYo_reaction}
\frac{\delta \ell_{\rm pi}}{\delta s}\bar D_t \varsigma =  - \sigma : \nabla u  + j _s \cdot \nabla D_t\gamma  + \sum_k j _k \cdot \nabla D_tw _k+\sum_a j_a D_t\upsilon_a, \;\; D_t \upsilon_a= \sum_k\bar\nu _{ak} D_t w_k
\end{equation}
and with respect to variations subject to $ \delta u = \partial _t \zeta+[ u , \zeta ] $ and to the \textit{variational constraint}
\begin{equation}\label{VC_GBYo_reaction}
\frac{\delta \ell_{\rm pi}}{\delta s}\bar D_\delta \varsigma = - \sigma : \nabla \zeta + j _s \cdot \nabla D_\delta\gamma  + \sum_kj _k \cdot \nabla D_\delta w _k + \sum_a j_a D_ \delta \upsilon_a, \;\; D_\delta \upsilon_a= \sum_k\bar\nu _{ak} D_\delta  w_k
\end{equation} 
with $ \delta w _k $, $ \delta \gamma $, and $ \zeta  $ vanishing at $t=0,T$.

A direct application of \eqref{Review_GBYo_reaction}--\eqref{VC_GBYo_reaction} yields the system
\begin{equation}\label{system_Eulerian_multi_reactions} 
\left\{
\begin{array}{l}
\vspace{0.2cm}\displaystyle
\partial _t \frac{\delta \ell_{\rm pi}}{\delta u}  + \pounds _ u \frac{\delta \ell_{\rm pi}}{\delta u} = \sum_k\rho _k \nabla \frac{\delta \ell_{\rm pi}}{\delta \rho _k }+ s \nabla \frac{\delta \ell_{\rm pi} }{\delta s }+\operatorname{div}  \sigma \\
\vspace{0.2cm}\displaystyle - \frac{\delta \ell_{\rm pi}}{\delta s } (\bar D_t s + \operatorname{div} j _s ) =  \sigma \!: \!\nabla u + j_s\! \cdot  \!\nabla \frac{\delta \ell_{\rm pi} }{\partial s } + \sum_kj _k \!\cdot \! \nabla \frac{\delta \ell_{\rm pi}}{\delta \rho _k }+ \sum_{a,k} j_a \bar\nu _{ak} \frac{\delta \ell_{\rm pi}}{\delta \rho  _k}\\
\displaystyle \bar D_ t\rho _k+  \operatorname{div}  j_k=\sum_a j_a \bar\nu _{ak},\;\;k=1,...,K,
\end{array}
\right.
\end{equation}
with boundary conditions $u|_{ \partial \Omega }=0$, see \cite{GBYo2017b}. In addition to the conditions \eqref{additional_conditions_multi}, the principle also yields the conditions
\begin{equation}\label{additional_conditions_multi_reaction} 
D_t \upsilon_a= -\sum_k\bar\nu_{ak}\frac{\delta \ell_{\rm pi} }{\delta \rho_k}, \quad a=1,...,r,
\end{equation}
consistently with the interpretation of $ \upsilon_a$ as being the thermodynamic displacement associated to chemical reaction $a$.
By using $\sum_kj_k=0$ and the Lavoisier law \eqref{Lavoisier_law}, the entropy equation in \eqref{system_Eulerian_multi_reactions} with Lagrangian $\ell_{\rm pi}$ in \eqref{ell_PI_multi}  takes the form
\[
\displaystyle T^*(\bar D_ts  + \operatorname{div} j _s) =  \sigma : \nabla u - j _s\cdot  \nabla T^* - \sum_k j _k\cdot \nabla \mu _k ^* + \sum_a j_a \mathcal{A} ^*_a\geq 0
\]
with modified chemical affinity given by
\[
\mathcal{A} _a^* = - \sum_k\bar\nu_{ak} \mu _k^*.
\]
As earlier, the variational formulation directly shows the occurrence of the modified thermodynamic forces $ \nabla T^*$, $ \nabla \mu _k^*$, and $ \mathcal{A} _a^*$ in the entropy production equation, which is crucial to achieve thermodynamic consistency. Thanks to this form of entropy production, the system can be closed by phenomenological relations as in the standard compressible multicomponent reacting fluid case, see \cite{dGMa1969}, by using the modified thermodynamic forces.

\subsection{Elliptic pressure equation}
The same procedure as used before (discovering hidden constraints associated with $\mathcal{C}$) can be applied for multicomponent semi-compressible fluids.

\subsubsection{Reversible Dynamics}
Consider the general GFD Lagrangian in enthalpy form for multicomponent semi-compressible fluids, which is
\begin{equation}
\ell(u,\{\rho_k\},s,p^\prime) = \tilde{\ell}(u,\{\rho_k\},s) + \mathcal{C}(\{\rho_k\},s,p^\prime)
\end{equation}
with
\begin{equation}
\mathcal{C}(\{\rho_k\},s,p^\prime) = \int_ \Omega  p^\prime \left(1 - \frac{\rho}{F(\phi,\eta,\{q _k\})} \right){\rm d}x
\end{equation}
for some arbitrary function $F(\phi,\eta,\{q _k\})$, recalling that $\eta = \frac{s}{\rho}$, $q_k = \frac{\rho_k}{\rho}$ and $\rho = \sum_k \rho_k$.

Following the same exact procedure as before (and omitting the details) we obtain
\begin{eqnarray}
\frac{\delta \ell}{\delta \rho  _k }  = \frac{\delta \tilde\ell}{\delta \rho  _k}  - \frac{p^\prime}{F} - \frac{p^\prime \eta}{F^2} \pp{F}{\eta} + \frac{p^\prime}{F^2} \pp{F}{q_k} - \sum_{k^\prime} \frac{p^\prime q_{k^\prime}}{F^2}  \pp{F}{q_{k^\prime}}, \qquad \frac{\delta \ell}{\delta s}  = \frac{\delta \tilde\ell}{\delta s}  + \frac{p^\prime}{F^2} \pp{F}{\eta}
\end{eqnarray}
Proceeding similarly with the single component case, we get the operators
\begin{equation}
\mathcal{G} p^\prime = \nabla \Big(\frac{p^\prime}{F}\Big) + \frac{p^\prime}{F^2} \pp{F}{\eta} \nabla \eta + \sum_k \frac{p^\prime}{F^2} \pp{F}{q_k} \nabla q_k =  \frac{1}{F} \nabla p^\prime - \frac{p^\prime}{F^2} \pp{F}{\phi} \nabla \phi
\end{equation}
and 
\begin{align*}
\mathcal{D} v &= \frac{1}{F} \operatorname{div}v + \frac{1}{F^2} \pp{F}{\phi}(v \cdot \nabla \phi)\\ &=\frac{1}{F^2}  \operatorname{div} ( Fv) - \frac{1}{F^2} \pp{F}{\eta}(v \cdot \nabla \eta)- \sum_k\frac{1}{F^2} \pp{F}{q_k}(v \cdot \nabla q_k).
\end{align*}
With these operators, the pressure Poisson equation for multicomponent semi-compressible models takes the form \eqref{Elliptlic_equation_reversible}, with $ \mathcal{E} $ given by
\begin{equation}\label{E_multi}
\mathcal{E} = \pounds _u \Big(\frac{1}{\rho} \frac{\delta \tilde\ell}{\delta u}\Big)  - \sum_k q_k \nabla \frac{\delta \tilde\ell}{\delta \rho  _k }  - \eta \nabla \frac{\delta \tilde\ell}{\delta s}.
\end{equation}
This is essentially the same equation as in the single component case, with the straightforward addition of terms involving $q_k$.

\paragraph{Anelastic Equations.} The operators $ \mathcal{D} $ and $ \mathcal{G} $ for the multicomponent anelastic/Boussinesq equations are in fact identical to those of the single component case, since $F(\phi,\eta,q_k) = \rho  _0(\phi)$. The expression $ \mathcal{E} = u \cdot \nabla u + 2 \omega \times u - b \nabla \phi $ found from \eqref{E_multi} also coincides with that of the single component case. This is also true when irreversible processes are introduced (see below).

\paragraph{Pseudoincompressible Equations.} For the pseudoincompressible equations we get the operators
\begin{align*}
\mathcal{G} f &=  \nabla\Big(\frac{f}{\rho^*}\Big) - f \Gamma^* \nabla \eta - \sum_k f \xi^*_k \nabla q_k  = \frac{1}{\rho^*}\Big( \nabla f - f \kappa _{\rm ad}^* \nabla p_0\Big)\\
\mathcal{D} v &= \frac{1}{ (\rho  ^* )^2}\operatorname{div}\big( \rho  ^* v\big) + \Gamma^* v \cdot \nabla \eta + \sum_k \xi^*_k v \cdot \nabla q_k=  \frac{1}{ \rho  ^* }\Big( \operatorname{div}v+ \kappa _{\rm ad}^* \nabla p_0 \cdot v\Big),
\end{align*}
where $\Gamma^* = \Gamma(p_0, \eta, \{q_k\}) = \frac{\partial^2 h}{\partial p \partial \eta} (  p_0 , \eta, \{q_k\})$, $ \kappa ^* _{\rm ad}= \kappa _{\rm ad}(p_0, \eta ,\{q_k\})= - \rho  ^*(p_0, \eta, \{q_k\}) \frac{\partial ^2 h}{\partial p^2}(p_0, \eta , \{q_k\})$, and $\xi^*_k = \frac{\partial^2 h}{\partial p \partial q_k} (p_0, \eta, \{q_k\})$, along with the same expression for $ \mathcal{E} $ as in the anelastic case.

A natural question that arises for the multicomponent pseudoincompressible equations is the existence of a simplified form for the operator $\mathcal{D} v$ for some special choice of thermodynamic potential, much like the case of a perfect ideal gas for the single component pseudoincompressible equations. This does in fact occur, for the ``constant $\kappa$" approximation for a moist gas (with or without condensates), where the simplified form is again $ \operatorname{div}(\rho_0 \theta_{v0}u ) = 0$. Here $\theta_v$ is an \textit{entropic variable} called the \textit{virtual potential temperature}. This arises because the constant $\kappa$ approximation has
\begin{equation}\label{k_constant_approx}
p \alpha = \kappa_d \theta_v \Pi_d
\end{equation}
and $ \kappa _{\rm ad}^* = \frac{1}{ p_0} \frac{C_{vd}}{C_{pd}}$, where $\Pi_d = C_{pd} (\frac{p}{p_{00}})^{\kappa_d}$ and $ \kappa _d= \frac{R_d}{C_{pd}}$. This is of the same form as for a single component perfect ideal gas, and therefore the same manipulations hold as before, which yields $ \mathcal{D} u= \operatorname{div}( \rho  _0^* \theta _{v0} u)\frac{1}{{\rho ^* }^2 \theta _v}$. More information about the thermodynamics of the constant $\kappa$ approximation can be found in \cite{Eldred2021}.

\subsubsection{Irreversible dynamics}

As for single component semi-compressible fluids, for general multicomponent semi-compressible fluids with irreversible processes the elliptic equation quickly becomes intractable due to the complicated dependence of thermodynamic parameterizations on $p^\prime$. However, for the multicomponent anelastic equations things again drastically simplify  and in fact we get the same elliptic equation as the single component anelastic equations.

\section{Conclusions}
\label{conclusions}

In this paper, we have presented the variational formulation of single and multicomponent semi-compressible models with irreversible processes: the Boussinesq, anelastic and pseudoincompressible equations, with arbitrary thermodynamic potentials and geopotentials.  We also gave evolution equations for a wide range of thermodynamic variables: $s$, $\eta$, $b$, $T$, $\theta$ and $\chi$. For the anelastic equations we have shown how in the case of an ideal gas various choices of reference profile lead to simplified $b$ (and other) expressions, connecting to existing anelastic equation sets in the geophysical and astrophysical fluid dynamics literature. Finally, we presented the elliptic pressure equation in the case of reversible dynamics for all models and for the Boussinesq/anelastic equations with irreversible dynamics; and highlighted the difficulties of formulating this equation for the pseudoincompressible equations with irreversible dynamics. Although developed in $\mathbb{R}^3$, the developments above are intrinsic and coordinate free; and therefore valid on arbitrary manifolds. 

Building on the foundation presented in this paper, a natural extension is to consider the Hamiltonian counterpart to Lagrangian variational formulations: Poisson and (metriplectic) bracket formulations for semi-compressible fluids. This will be the subject of future work, utilizing the same approach as in \cite{ElGB2018}.

\section{Acknowledgements}

This research was supported as part of the Energy Exascale Earth
System Model (E3SM) project, funded by the U.S. Department of Energy, Office of Science, Office of Biological and Environmental Research.

This research was supported by the Exascale Computing Project (17?SC?20?SC), a collaborative effort of the U.S. Department of Energy Office of Science and the National Nuclear Security Administration. 

Christopher Eldred was funded for part of this work by French National Research Agency through contract ANR-14-CE23-0010 (HEAT) while at Inria Grenoble Rhone-Alpes. 

Sandia National Laboratories is a multimission laboratory managed and operated by National Technology and Engineering Solutions of Sandia, LLC, a wholly owned subsidiary of Honeywell International Inc., for the U.S. Department of Energy's National Nuclear Security Administration under Contract DE-NA0003525. This paper describes objective technical results and analysis. Any subjective views or opinions that might be expressed in the paper do not necessarily represent the views of the U.S. Department of Energy or the United States Government.

\appendix
\section{Derivation of the multicomponent pseudoincompressible model}\label{appendix_A}

We show how the general Lagrangian system \eqref{system_Eulerian_atmosphere_moist_L} produces the multicomponent pseudoincompressible model \eqref{Eulerian_PI_multi} when the Lagrangian function \eqref{ell_PI_multi} is used. The functional derivatives are given in \eqref{functional_der_pi}. From the conditions \eqref{mass_conservation_pi}, the total mass equation is $\bar D_t \rho  =0$, from which the fluid momentum equation in  \eqref{Eulerian_PI_multi} can be rewritten as
\[
(\partial _t + \pounds _u) \Big(\frac{1}{\rho} \frac{\delta \ell}{\delta u}\Big)  = \sum_k q_k \nabla \frac{\delta \ell}{\delta \rho  _k }  + \eta \nabla \frac{\delta \ell}{\delta s} + \frac{1}{ \rho  } \operatorname{div} \sigma.
\]
Using \eqref{functional_der_pi}, we get
\begin{align*}
&\partial _t u + u \cdot \nabla (u+R) + \nabla u^\mathsf{T} \cdot (u+R)= \nabla \left( \frac{1}{2} |u| ^2 + R \cdot u \right)\\
& \qquad \qquad \qquad - \sum_k q _k \nabla \left( \frac{\partial h}{\partial q_k} + \phi + p' \frac{\partial ^2 h}{\partial p \partial q _k } \right)  - \eta \nabla \left( \frac{\partial h}{\partial \eta }+ p' \frac{\partial ^2 h}{\partial p \partial \eta }  \right)  
\end{align*}
which yields
\[
\partial _t u + u \cdot \nabla u + 2 \omega \times u =- \nabla \phi - \sum_k q _k \nabla \left( \frac{\partial h}{\partial q_k} + p' \frac{\partial ^2 h}{\partial p \partial q _k } \right) - \eta \nabla \left( \frac{\partial h}{\partial \eta }+ p' \frac{\partial ^2 h}{\partial p \partial \eta }  \right).
\]
By elementary computations, the terms involving $h$ can be simplified as 
\[
\frac{\partial h}{\partial p} \nabla p_0 + \frac{\partial h}{\partial p} \nabla p' + p' \frac{\partial ^2 h}{\partial p ^2  } \nabla p_0 = - \frac{ \rho  _0}{ \rho  } \nabla \phi + \frac{1}{ \rho  } \nabla p' - \frac{1}{\rho} p' \kappa _{\rm ad} \nabla p_0,
\]
where we recall that $h$, $ \kappa _{\rm ad}$, and $ \rho  $ are evaluated at $(p_0, \eta , \{q_k\})$. We thus obtain the momentum equation in \eqref{Eulerian_PI_multi}. The single component case in \eqref{Eulerian_PI} follows similarly.

{
\footnotesize

\bibliographystyle{new}

\end{document}